%% file: 331T7final.tex
\documentclass[preprintnumbers]{revtex4}
\usepackage{amsfonts}
\usepackage{amsmath}
\usepackage{hyperref}
\usepackage{amssymb}
\usepackage[english]{babel}
\usepackage{graphicx}
\usepackage{epsfig}
\usepackage{bm}
\usepackage{longtable}
\usepackage{verbatim}
\usepackage{longtable}

\newcommand{\mathsym}[1]{{}}

\input{tcilatex}
\begin{document}

\title{Fermion mass and mixing pattern in a minimal T7 flavor 331 model.}
\author{A. E. C\'arcamo Hern\'andez${}^{a}$}
\email{antonio.carcamo@usm.cl}
\author{R. Martinez${}^{{b}}$}
\email{remartinezm@unal.edu.co}
\affiliation{$^{{a}}$Universidad T\'{e}cnica Federico Santa Mar\'{\i}a\\
and Centro Cient\'{\i}fico-Tecnol\'{o}gico de Valpara\'{\i}so\\
Casilla 110-V, Valpara\'{\i}so, Chile,\\
$^{{b}}$Universidad Nacional de Colombia, Departamento de F\'{\i}sica,
Ciudad Universitaria, Bogot\'{a} D.C., Colombia. }
\date{\today }

\begin{abstract}
We present a model based on the $SU(3)_{C}\otimes SU(3)_{L}\otimes U(1)_{X}$
gauge symmetry having an extra $T_{7}\otimes Z_{2}\otimes Z_{3}\otimes
Z_{14} $ flavor group, which successfully describes the observed SM fermion
mass and mixing pattern. In this framework, the light active neutrino masses
arise via double seesaw mechanism and the observed charged fermion mass and
quark mixing hierarchy is a consequence of the $Z_{2}\otimes Z_{3}\otimes
Z_{14}$ symmetry breaking at very high energy. In our minimal $T_{7}$ flavor
331 model, the spectrum of neutrinos includes very light active neutrinos as
well as heavy and very heavy sterile neutrinos. The model has in total 16
effective free parameters, which are fitted to reproduce the experimental
values of the 18 physical observables in the quark and lepton sectors. The
obtained physical observables for both quark and lepton sectors are
compatible with their experimental values. The model predicts the effective
Majorana neutrino mass parameter of neutrinoless double beta decay to be $%
m_{\beta \beta }=$ 3 and 40 meV for the normal and the inverted neutrino
spectrum, respectively. Furthermore, our model features a vanishing leptonic
Dirac CP violating phase.
\end{abstract}

\maketitle

\input{convention.tex}



\section{Introduction}

In spite of the experimental confirmation of the big accomplishments 
of the Standard Model (SM) in describing electroweak phenomena, given by the
discovery of the $\sim $ $126$ GeV Higgs boson by ATLAS and CMS
collaborations at the CERN Large Hadron Collider (LHC) \cite%
{atlashiggs,cmshiggs,newtevatron,CMS-PAS-HIG-12-020}, there are several
issues that the SM is unable to address. 
Some of these issues are the Dark Matter problem, the fermion mass and
mixing hierarchy and the neutrino oscillations \cite{SM,PDG}. Furthermore,
there are problems with the matter-antimatter asymmetry related with new
phases responsible for CP violation. Moreover, the SM does not explain the
tiny value of the neutron dipole moment. Because of these reasons it is
necesary to consider an extension of the Standard Model. In particular, the
observed the quark mass and mixing pattern, which cannot be explained from
first principles in the context of the Standard Model, is a clear indication
of physics beyond the Standard Model. On the other hand, the discovery of
the Higgs boson opens the possibility to unravel the Electroweak Symmetry
Breaking (EWSB) mechanism and motivates to study extensions of the SM having
extra scalar particles that could provide an explanation for the existence
of Dark Matter \cite{BSMtheorieswithDM}.

\quad The lack of predictivity of the Standard Model Yukawa sector,
motivates to consider extensions of the Standard Model aimed to address its
flavor puzzle. Discrete flavor symmetries are important because they
generate fermion textures useful to explain the three generation flavor
structure
, for recent reviews see Refs. \cite%
{King:2013eh,Altarelli:2010gt,Ishimori:2010au}. These discrete flavour
symmetries have been employed in extensions of the Standard Model with the
aim to study the fermion mass and mixing hierarchy in order to address the
flavor puzzle of the SM. Discrete flavor symmetries can arise from the
underlying theory, e.g., string theory or compactification via orbifolding.
In particular, from heterotic orbifold models, 
one can generate the $D_{4}$ and $\Delta(54)$ flavor symmetries \cite%
{StringsandDS}. Furthermore, magnetized/intersecting D-brane models can
generate the $\Delta(27)$ flavor symmetry \cite{StringsandDS}. Discrete
symmetries may link the low energy physics and the underlying theory.

\quad Furthermore, another unaswered issue in particle physics is the
existence of three generations of fermions at low energies. The mixing
patterns of leptons and quarks are significantly different; while in the
quark sector, the mixing angles are small, in the lepton sector two of the
mixing angles are large and one is small. Models having $SU(3)_{C}\otimes
SU(3)_{L}\otimes U(1)_{X}$ as a gauge symmetry, are vectorlike with three
fermion generations and thus do not contain anomalies \cite%
{331-pisano,331-frampton,331-long,M-O,anomalias}. Defining the electric
charge as the linear combination of the $T_{3}$ and $T_{8}$ $SU(3)_{L}$
generators, we have that it is a free parameter, which does not depend on
the anomalies ($\beta $). The charge of the exotic particles is defined by
setting a value for the $\beta $ parameter. Setting $\beta =-\frac{1}{\sqrt{3%
}}$, implies that the third component of the weak lepton triplet is a
neutral field $\nu _{R}^{C}$ allowing to build the Dirac Yukawa term with
the usual field $\nu _{L}$ of the weak doublet. By adding very heavy sterile
neutrinos $N_{R}^{1,2,3}$ in the model, light neutrino masses can be
generated via double seesaw mechanism. The 331 models with $\beta =-\frac{1}{%
\sqrt{3}}$ provide an alternative neutrino mass generation mechanism and
include in their neutrino spectrum light active sub-eV scale neutrinos as
well as sterile neutrinos which could be dark matter candidates if they are
light enough or candidates for detection at the LHC, if they have TeV scale
masses. Having TeV scale sterile neutrinos in its neutrino spectrum, makes
the 331 models very important since if these sterile neutrinos are detected
at the LHC, these models can shed light in the understanding of the
electroweak symmetry breaking mechanism.

\quad Neutrino oscillation experiments \cite%
{PDG,An:2012eh,Abe:2011sj,Adamson:2011qu,Abe:2011fz,Ahn:2012nd} show that
there are at most one massless active neutrino and that the different
neutrino flavors mix. Neutrino oscillations experiments do not determine
neither the absolute value of the neutrino masses nor the Majorana or Dirac
feature of the neutrino. Nevertheless neutrino mass bounds can be obtained
from tritio beta decay \cite{Kraus:2004zw}, double beta decay \cite%
{Doublebetadecay} and cosmology \cite{Ade:2013zuv}.

\quad The global fits of the available data from the Daya Bay \cite%
{An:2012eh}, T2K \cite{Abe:2011sj}, MINOS \cite{Adamson:2011qu}, Double
CHOOZ \cite{Abe:2011fz} and RENO \cite{Ahn:2012nd} neutrino oscillation
experiments, constrain the neutrino mass squared splittings and mixing
parameters \cite{Forero:2014bxa}. The current experimental data on neutrino
mixing parameters suggests a violation of the tribimaximal symmetry
described by the Tribimaximal Mixing (TBM) matrix, 
whose predicted mixing angles satisfy $\left( \sin ^{2}\theta _{12}\right)
_{TBM}=\frac{1}{3}$, $\left( \sin ^{2}\theta _{23}\right) _{TBM}=\frac{1}{2}$%
, and $\left( \sin ^{2}\theta _{13}\right) _{TBM}=0$. To generate nearly
tribimaximal leptonic mixing angles consistent with the experimental data,
discrete symmetry groups \cite%
{discrete-lepton,discrete-quark,s3pheno,331S3leptons,331S3quarks,2HDMS3,warpedS3,Delta27,Tprime,T7}
are implemented in extensions of the Standard Model. Another approach to
address the flavor puzzle consists in postulating fermion mass textures (see
Ref \cite{textures} for some works considering textures). Moreover, models
based on extended symmetries in the context of Multi-Higgs sectors, Grand
Unification, Extradimensions and Superstrings have been explored \cite%
{King:2013eh,GUT,Extradim,String,horizontal} to provide an explanation for
the observed fermion mass and mixing pattern. Furthermore, in the framework
of minimal 331 models, the discrete groups require of an extra high energy
scale, larger than the scale of breaking of the $SU(3)_{C}\otimes
SU(3)_{L}\otimes U(1)_{X}$ symmetry, and in some cases, new scalar fields
need to be introduced at the very high discrete symmetry breaking scale,
with the aim to fulfill the irreducible representations of these discrete
groups that allow to get viable fermion textures at low energies, after the
gauge and discrete symmetries are spontaneosly broken.

\quad In this paper we formulate an extension of the minimal $%
SU(3)_{C}\times SU(3)_{L}\times U(1)_{X}$ model with $\beta =-\frac{1}{\sqrt{%
3}}$, where an extra \mbox{$T_{7}\otimes Z_{2}\otimes Z_{3}\otimes Z_{14}$}
discrete group (see Ref \cite{T7} for studies about the $T_{7}$ flavor
group) extends the symmetry of the model and very heavy extra scalar fields
are added with the aim to generate viable and predictive textures for the
fermion sector that successfully describe the fermion mass and mixing
pattern. The obtained physical observables in the quark and lepton sector
are consistent with the experimental data. Our model at low energies reduces
to the minimal $331$ model with $\beta =-\frac{1}{\sqrt{3}}$.

\quad The content of this paper goes as follows. In Sec. \ref{model} we
describe the proposed model. 
Sec. \ref{leptonmassesandmixing} is devoted to the discussion of lepton
masses and mixings. In Sec. \ref{quarkmassesandmixing}, we present our
results in terms of quark masses and mixing, which is followed by a
numerical analysis. Conclusions are stated in Sec. \ref{conclusions}.
Appendix \ref{ap1} provides a brief description of the $T_{7}$ discrete
group.


\section{The model}

\label{model}

\subsection{Particle content}

We consider a $SU(3)_{C}\otimes SU(3)_{L}\otimes U(1)_{X}\otimes
T_{7}\otimes Z_{2}\otimes Z_{3}\otimes Z_{14}$ model where the full symmetry 
$\mathcal{G}$ is spontaneously broken in three steps as follows: 
\begin{eqnarray}
&&\mathcal{G}=SU(3)_{C}\otimes SU\left( 3\right) _{L}\otimes U\left(
1\right) _{X}\otimes T_{7}\otimes Z_{2}\otimes Z_{3}\otimes Z_{14}{%
\xrightarrow{\Lambda _{int}}}  \notag \\
&&\hspace{7mm}SU(3)_{C}\otimes SU\left( 3\right) _{L}\otimes U\left(
1\right) _{X}{\xrightarrow{v_{\chi }}}SU(3)_{C}\otimes SU\left( 2\right)
_{L}\otimes U\left( 1\right) _{Y}{\xrightarrow{v_{\eta },v_{\rho }}}  \notag
\\
&&\hspace{7mm}SU(3)_{C}\otimes U\left( 1\right) _{Q},  \label{Group}
\end{eqnarray}%
where the hierarchy $v_{\eta },v_{\rho }\ll v_{\chi }\ll \Lambda _{int}$
among the symmetry breaking scales is fullfilled.


\quad The electric charge in our 331 model is defined as: 
\begin{equation}
Q=T_{3}-\frac{1}{\sqrt{3}}T_{8}+XI,
\end{equation}%
where $T_3$ and $T_8$ are the $SU(3)_L$ diagonal generators and $I$ is the $%
3\times 3$ identity matrix. 

\quad Two families of quarks are grouped in a $3^{\ast }$ irreducible
representations (irreps), as required from the $SU(3)_{L}$ anomaly
cancellation. Furthermore, from the quark colors, it follows that the number
of $3^{\ast }$ irreducible representations is six. The other family of
quarks is grouped in a $3$ irreducible representation. Moreover, there are
six $3$ irreps taking into account the three families of leptons.
Consequently, the $SU(3)_{L}$ representations are vector like and do not
contain anomalies. The quantum numbers for the fermion families are assigned
in such a way that the combination of the $U(1)_{X}$ representations with
other gauge sectors is anomaly free. Therefore, the anomaly cancellation
requirement implies that quarks are unified in the following $%
(SU(3)_{C},SU(3)_{L},U(1)_{X})$ left- and right-handed representations: 
\begin{align}
Q_{L}^{1,2}& =%
\begin{pmatrix}
D^{1,2} \\ 
-U^{1,2} \\ 
J^{1,2} \\ 
\end{pmatrix}%
_{L}:(3,3^{\ast },0),\hspace{1cm}Q_{L}^{3}=%
\begin{pmatrix}
U^{3} \\ 
D^{3} \\ 
T \\ 
\end{pmatrix}%
_{L}:(3,3,1/3),  \label{fermion_spectrumleft} \\
& 
\begin{array}{c}
D_{R}^{1,2,3}:(3,1,-1/3), \\ 
J_{R}^{1,2}:(3,1,-1/3),%
\end{array}%
\hspace{0.7cm}%
\begin{array}{c}
U_{R}^{1,2,3}:(3,1,2/3), \\ 
T_{R}:(3,1,2/3).%
\end{array}
\label{fermion_spectrumright}
\end{align}%
Here $U_{L}^{i}$ and $D_{L}^{i}$ ($i=1,2,3$) are the left handed up- and
down-type quarks in the flavor basis. The right handed SM\ quarks $U_{R}^{i}$
and $D_{R}^{i}$ ($i=1,2,3$) and right handed exotic quarks $T_{R}$ and $%
J_{R}^{1,2}$ are assigned into $SU(3)_{L}$ singlets representations, so that
their $U(1)_{X}$ quantum numbers correspond to their electric charges.

Furthermore, cancellation of anomalies implies that leptons are grouped in
the following $(SU(3)_{C},SU(3)_{L},U(1)_{X})$ left- and right-handed
representations: 
\begin{align}
L_{L}^{1,2,3}& =%
\begin{pmatrix}
\nu ^{1,2,3} \\ 
e^{1,2,3} \\ 
(\nu ^{1,2,3})^{c} \\ 
\end{pmatrix}%
_{L}:(1,3,-1/3), \\
& 
\begin{array}{c}
e_{R}:(1,1,-1), \\ 
N_{R}^{1}:(1,1,0), \\ 
\end{array}%
\hspace{0.7cm}%
\begin{array}{c}
\mu _{R}:(1,1,-1), \\ 
N_{R}^{2}:(1,1,0), \\ 
\end{array}%
\hspace{0.7cm}%
\begin{array}{c}
\tau _{R}:(1,1,-1), \\ 
N_{R}^{3}:(1,1,0). \\ 
\end{array}%
\end{align}
where $\nu _{L}^{i}$ and $e_{L}^{i}$ ($e_{L},\mu _{L},\tau _{L}$) are the
neutral and charged lepton families, respectively. Let's note that we assign
the right-handed leptons as $SU(3)_{L}$ singlets, which implies that their $%
U(1)_{X}$ quantum numbers correspond to their electric charges. The exotic
leptons of the model are: three neutral Majorana leptons $(\nu
^{1,2,3})_{L}^{c}$ and three right-handed Majorana leptons $N_{R}^{1,2,3}$
(A recent discussion of double and inverse see-saw neutrino mass generation
mechanisms in the context of 331 models can be found in Ref. \cite{catano}).

\quad The scalar sector the 331 models includes: three $3$'s irreps of $%
SU(3)_{L}$, where one triplet $\chi $ gets a TeV scale vaccuum expectation
value (VEV) $v_{\chi }$, that breaks the $SU(3)_{L}\times U(1)_{X}$ symmetry
down to $SU(2)_{L}\times U(1)_{Y}$, thus generating the masses of non SM
fermions and non SM gauge bosons; and two light triplets $\eta $ and $\rho $
acquiring electroweak scale VEVs $v_{\eta }$ and $v_{\rho }$, respectively
and thus providing masses for the fermions and gauge bosons of the SM.

\quad Regarding the scalar sector of the minimal 331 model, we assign the
scalar fields in the following $[SU(3)_{L},U(1)_{X}]$ representations: 
\begin{align}
\chi & =%
\begin{pmatrix}
\chi _{1}^{0} \\ 
\chi _{2}^{-} \\ 
\frac{1}{\sqrt{2}}(\upsilon _{\chi }+\xi _{\chi }\pm i\zeta _{\chi }) \\ 
\end{pmatrix}%
:(3,-1/3),\hspace{1cm}\rho =%
\begin{pmatrix}
\rho _{1}^{+} \\ 
\frac{1}{\sqrt{2}}(\upsilon _{\rho }+\xi _{\rho }\pm i\zeta _{\rho }) \\ 
\rho _{3}^{+} \\ 
\end{pmatrix}%
:(3,2/3),  \notag \\
\eta & =%
\begin{pmatrix}
\frac{1}{\sqrt{2}}(\upsilon _{\eta }+\xi _{\eta }\pm i\zeta _{\eta }) \\ 
\eta _{2}^{-} \\ 
\eta _{3}^{0}%
\end{pmatrix}%
:(3,-1/3).  \label{331-scalar}
\end{align}

We extend the scalar sector of the minimal 331 model by adding the following
eleven very heavy $SU(3)_{L}$ scalar singlets: 
\begin{equation}
\sigma \sim (1,0),\hspace{1cm}\tau \sim (1,0),\hspace{1cm}\xi _{j}:(1,0),%
\hspace{1cm}\zeta _{j}:(1,0),\hspace{1cm}S_{j}:(1,0),\hspace{1cm}j=1,2,3.
\label{331scalarsextra}
\end{equation}

We assign the scalars into $T_{7}$ triplet, $T_{7}$ antitriplet and $T_{7}$
singlet representions. The $T_{7}\otimes Z_{2}\otimes Z_{3}\otimes Z_{14}$
assignments of the scalar fields are: 
\begin{eqnarray}
\eta &\sim &\left( \mathbf{1}_{0},1\mathbf{,}e^{\frac{2\pi i}{3}},1\right) ,%
\hspace{1cm}\rho \sim \left( \mathbf{1}_{0}\mathbf{,}1,e^{-\frac{2\pi i}{3}%
},1\right) ,\hspace{1cm}\chi \sim \left( \mathbf{1}_{0}\mathbf{,}%
1,1,1\right) ,\hspace{1cm}\tau \sim \left( \mathbf{1}_{1}\mathbf{,-}%
1,1,1\right) ,\hspace{1cm}\hspace{1cm}  \notag \\
\xi &\sim &\left( \mathbf{3,}1,e^{\frac{2\pi i}{3}},1\right) ,\hspace{1cm}%
\zeta \sim \left( \overline{\mathbf{3}}\mathbf{,}1,1,1\right) ,\hspace{1cm}%
S\sim \left( \mathbf{3,}1,e^{\frac{2\pi i}{3}},1\right) ,\hspace{1cm}\sigma
\sim \left( \mathbf{1}_{0}\mathbf{,}1,1,e^{-\frac{i\pi }{7}}\right) .
\end{eqnarray}

It is noteworthy that the $SU(3)_{L}$ singlet scalar field $\tau $ is the
only scalar odd under the $Z_{2}$ symmetry and assigned as a non trivial $%
T_{7}$ singlet. This scalar field $\tau $ is crucial for explaining the
hierarchy between the SM up and SM down type quark masses.

\quad In the concerning to the lepton sector, we have the following $%
T_{7}\otimes Z_{2}\otimes Z_{3}\otimes Z_{14}$ assignments: 
\begin{eqnarray}
L_{L} &\sim &\left( \mathbf{3,}1,e^{-\frac{2\pi i}{3}},1\right) ,\hspace{1cm}%
e_{R}\sim \left( \mathbf{1}_{0}\mathbf{,}1,e^{-\frac{2\pi i}{3}},-1\right) ,%
\hspace{1cm}\mu _{R}\sim \left( \mathbf{1}_{1}\mathbf{,}1,e^{-\frac{2\pi i}{3%
}},e^{\frac{4i\pi }{7}}\right) ,  \notag \\
\tau _{R} &\sim &\left( \mathbf{1}_{2}\mathbf{,}1,e^{-\frac{2\pi i}{3}},e^{%
\frac{2i\pi }{7}}\right) ,\hspace{1cm}N_{R}\sim \left( \mathbf{3,}1,e^{-%
\frac{2\pi i}{3}},1\right) ,
\end{eqnarray}%
while the $T_{7}\otimes Z_{2}\otimes Z_{3}\otimes Z_{14}$ assignments for
the quark sector are: 
\begin{eqnarray}
Q_{L}^{1} &\sim &\left( \mathbf{1}_{0}\mathbf{,}1,1,e^{\frac{2\pi i}{7}%
}\right) ,\hspace{1cm}Q_{L}^{2}\sim \left( \mathbf{1}_{0}\mathbf{,}1,1,e^{%
\frac{\pi i}{7}}\right) ,\hspace{1cm}Q_{L}^{3}\sim \left( \mathbf{1}_{0}%
\mathbf{,}1,1,1\right) ,  \notag \\
U_{R}^{1} &\sim &\left( \mathbf{1}_{0}\mathbf{,}1,e^{-\frac{2\pi i}{3}},e^{%
\frac{2\pi i}{7}}\right) ,\hspace{1cm}U_{R}^{2}\sim \left( \mathbf{1}_{0}%
\mathbf{,}1,e^{-\frac{2\pi i}{3}},e^{\frac{\pi i}{7}}\right) ,\hspace{1cm}%
U_{R}^{3}\sim \left( \mathbf{1}_{0}\mathbf{,}1,e^{-\frac{2\pi i}{3}%
},1\right) ,  \notag \\
D_{R}^{1} &\sim &\left( \mathbf{1}_{0}\mathbf{,-}1,e^{\frac{2\pi i}{3}},e^{%
\frac{2\pi i}{7}}\right) ,\hspace{1cm}D_{R}^{2}\sim \left( \mathbf{1}_{0}%
\mathbf{,-}1,e^{\frac{2\pi i}{3}},e^{\frac{\pi i}{7}}\right) ,\hspace{1cm}%
D_{R}^{3}\sim \left( \mathbf{1}_{0}\mathbf{,-}1,e^{\frac{2\pi i}{3}%
},1\right) ,  \notag \\
T_{R} &\sim &\left( \mathbf{1}_{0}\mathbf{,}1,1,1\right) ,\hspace{1cm}%
J_{R}^{1}\sim \left( \mathbf{1}_{0}\mathbf{,}1,1,e^{\frac{2\pi i}{7}}\right)
,\hspace{1cm}J_{R}^{2}\sim \left( \mathbf{1}_{0}\mathbf{,}1,1,e^{\frac{\pi i%
}{7}}\right) .
\end{eqnarray}%
Here the dimensions of the $T_{7}$ irreducible representations are specified
by the numbers in boldface. 
In the concerning to the lepton sector, we recall that the left and right
handed leptons are grouped into $T_{7}$ triplet and $T_{7}$ singlet
irreducible representations, respectively, whereas the right handed Majorana
neutrinos are unified into a $T_{7}$ triplet. Regarding the quark sector, we
assign the quarks fields into trivial $T_{7}$ singlet representations. Note
that SM right handed quarks are the only quark fields transforming non
trivially under the $Z_{3}$ symmetry. Besides that, let's note that the
right handed SM down type quarks are the only fermions odd under the $Z_{2}$
symmetry. Furthermore, it is worth mentioning that the $SU(3)_{L}$ scalar
triplets are assigned to a $T_{7}$ trivial singlet representation, whereas
the $SU(3)_{L}$ scalar singlets are accomodated into two $T_{7}$ triplets,
one $T_{7}$ antitriplet, one $T_{7}$ trivial singlet and one $T_{7}$ non
trivial singlet. The $SU(3)_{L}$ scalar singlets $T_{7}$ triplets are
distinguished by their $Z_{3}$ charge assignments.

\quad With the aforementioned field content of our model, the relevant quark
and lepton Yukawa terms invariant under the group $\mathcal{G}$, take the
form: 
\begin{eqnarray}
-\mathcal{L}_{Y}^{\left( Q\right) } &=&y_{11}^{\left( U\right) }\overline{Q}%
_{L}^{1}\rho ^{\ast }U_{R}^{1}\frac{\sigma ^{4}}{\Lambda ^{4}}%
+y_{12}^{\left( U\right) }\overline{Q}_{L}^{1}\rho ^{\ast }U_{R}^{2}\frac{%
\sigma ^{3}}{\Lambda ^{3}}+y_{21}^{\left( U\right) }\overline{Q}_{L}^{2}\rho
^{\ast }U_{R}^{1}\frac{\sigma ^{3}}{\Lambda ^{3}}+y_{22}^{\left( U\right) }%
\overline{Q}_{L}^{2}\rho ^{\ast }U_{R}^{2}\frac{\sigma ^{2}}{\Lambda ^{2}} 
\notag \\
&&+y_{13}^{\left( U\right) }\overline{Q}_{L}^{1}\rho ^{\ast }U_{R}^{3}\frac{%
\sigma ^{2}}{\Lambda ^{2}}+y_{31}^{\left( U\right) }\overline{Q}_{L}^{3}\eta
U_{R}^{1}\frac{\sigma ^{2}}{\Lambda ^{2}}+y_{23}^{\left( U\right) }\overline{%
Q}_{L}^{2}\rho ^{\ast }U_{R}^{3}\frac{\sigma }{\Lambda }+y_{32}^{\left(
U\right) }\overline{Q}_{L}^{3}\eta U_{R}^{2}\frac{\sigma }{\Lambda }  \notag
\\
&&+y_{33}^{\left( U\right) }\overline{Q}_{L}^{3}\eta U_{R}^{3}+y^{\left(
T\right) }\overline{Q}_{L}^{3}\chi T_{R}+y_{1}^{\left( J\right) }\overline{Q}%
_{L}^{1}\chi ^{\ast }J_{R}^{1}+y_{2}^{\left( J\right) }\overline{Q}%
_{L}^{2}\chi ^{\ast }J_{R}^{2}+y_{33}^{\left( D\right) }\overline{Q}%
_{L}^{3}\rho D_{R}^{3}\frac{\tau ^{3}}{\Lambda ^{3}}  \notag \\
&&+y_{11}^{\left( D\right) }\overline{Q}_{L}^{1}\eta ^{\ast }D_{R}^{1}\frac{%
\sigma ^{4}\tau ^{3}}{\Lambda ^{7}}+y_{12}^{\left( D\right) }\overline{Q}%
_{L}^{1}\eta ^{\ast }D_{R}^{2}\frac{\sigma ^{3}\tau ^{3}}{\Lambda ^{6}}%
+y_{21}^{\left( D\right) }\overline{Q}_{L}^{2}\eta ^{\ast }D_{R}^{1}\frac{%
\sigma ^{3}\tau ^{3}}{\Lambda ^{6}}+y_{22}^{\left( D\right) }\overline{Q}%
_{L}^{2}\eta ^{\ast }D_{R}^{2}\frac{\sigma ^{2}\tau ^{3}}{\Lambda ^{5}} 
\notag \\
&&+y_{13}^{\left( D\right) }\overline{Q}_{L}^{1}\eta ^{\ast }D_{R}^{3}\frac{%
\sigma ^{2}\tau ^{3}}{\Lambda ^{5}}+y_{31}^{\left( D\right) }\overline{Q}%
_{L}^{3}\rho D_{R}^{1}\frac{\sigma ^{2}\tau ^{3}}{\Lambda ^{5}}%
+y_{23}^{\left( D\right) }\overline{Q}_{L}^{2}\eta ^{\ast }D_{R}^{3}\frac{%
\sigma \tau ^{3}}{\Lambda }+y_{32}^{\left( D\right) }\overline{Q}%
_{L}^{3}\rho D_{R}^{2}\frac{\sigma \tau ^{3}}{\Lambda }+H.c,
\label{Yukawaterms}
\end{eqnarray}%
\begin{eqnarray}
-\mathcal{L}_{Y}^{\left( L\right) } &=&h_{\rho e}^{\left( L\right) }\left( 
\overline{L}_{L}\rho \xi \right) _{\mathbf{\mathbf{1}}_{0}}e_{R}\frac{\sigma
^{7}}{\Lambda ^{8}}+h_{\rho \mu }^{\left( L\right) }\left( \overline{L}%
_{L}\rho \xi \right) _{\mathbf{1}_{2}}\mu _{R}\frac{\sigma ^{4}}{\Lambda ^{5}%
}+h_{\rho \tau }^{\left( L\right) }\left( \overline{L}_{L}\rho \xi \right) _{%
\mathbf{1}_{1}}\tau _{R}\frac{\sigma ^{2}}{\Lambda ^{3}}  \notag \\
&&+h_{\chi }^{\left( L\right) }\left( \overline{L}_{L}\chi N_{R}\right) _{%
\mathbf{\mathbf{1}}_{0}}+\frac{1}{2}h_{1N}\left( \overline{N}%
_{R}N_{R}^{C}\right) _{\mathbf{3}}\xi +h_{2N}\left( \overline{N}%
_{R}N_{R}^{C}\right) _{\overline{\mathbf{3}}}S  \notag \\
&&+h_{\rho }\varepsilon _{abc}\left( \overline{L}_{L}^{a}\left(
L_{L}^{C}\right) ^{b}\right) _{\mathbf{3}}\left( \rho ^{\ast }\right) ^{c}%
\frac{\zeta }{\Lambda }+H.c,  \label{Lylepton}
\end{eqnarray}%
where $y_{ij}^{\left( U,D\right) }$ ($i,j=1,2,3$), $h_{\rho e}^{\left(
L\right) }$, $h_{\rho \mu }^{\left( L\right) }$, $h_{\rho \tau }^{\left(
L\right) }$, $h_{\chi }^{\left( L\right) }$, $h_{1N}$, $h_{2N}$ and $h_{\rho
}$ are $\mathcal{O}(1)$ dimensionless couplings.

\quad In the following we explain the role each discrete group factors of
our model. The $T_{7}$ and $Z_{3}$ discrete groups reduce the number of the $%
SU(3)_{C}\otimes SU(3)_{L}\otimes U(1)_{X}$ model parameters. This allow us
to get predictive and viable textures for the fermion sector that
successfully describe the prevailing pattern of fermion masses and mixings,
as we will show in sections \ref{leptonmassesandmixing} and \ref%
{quarkmassesandmixing}. We use $T_{7}$ since it is the minimal non-Abelian
discrete group having a complex triplet \cite{Ishimori:2010au}, where the
three fermion generations can be naturally unified. The $Z_{3}$ symmetry
determines the allowed entries of the neutrino mass matrix. Furthermore, the 
$Z_{3}$ symmetry distinguishes the right handed exotic quaks, being neutral
under $Z_{3}$ from the right handed SM quarks, charged under this symmetry.
This results in the absence of mixing between SM quarks and exotic quarks.
Consequently, the $Z_{3}$ symmetry is crucial for decoupling the SM quarks
from the exotic quarks. The $Z_{2}$ symmetry separates the right handed SM
down type quarks, odd under this symmetry, from the SM up type quarks, even
under $Z_{2}$. Consequently, the $Z_{2}$ symmetry is responsible for the
mass hierarchy between SM up and SM down type quarks. Note that the $%
SU(3)_{L}$ scalar singlet $\tau $, the only $T_{7}$ non trivial singlet and $%
Z_{2}$ odd scalar, only appears in the SM down type quark Yukawa terms. The $%
Z_{14}$ symmetry generates the hierarchy among charged fermion masses and
quark mixing angles that yields the observed charged fermion mass and quark
mixing pattern. Furthermore, it is noteworthy that the five dimensional
Yukawa operators $\frac{1}{\Lambda }\left( \overline{L}_{L}\rho \xi \right)_{%
\mathbf{\mathbf{1}}_{0}}e_{R}$, $\frac{1}{\Lambda }\left( \overline{L}%
_{L}\rho \xi \right) _{\mathbf{\mathbf{1}}_{2}}\mu _{R}$ and $\frac{1}{%
\Lambda }\left( \overline{L}_{L}\rho \xi \right) _{\mathbf{\mathbf{1}}%
_{1}}\tau _{R}$ are invariant under $T_{7}$ but not under the $Z_{14}$
symmetry, because the right handed charged leptons are $Z_{14}$ charged. We
use $Z_{14}$ because it is the smallest lowest cyclic symmetry, that allows
to build a twelve dimensional charged lepton Yukawa term, from a $\frac{%
\sigma ^{7}}{\Lambda ^{7}}$ insertion on the $\frac{1}{\Lambda }\left( 
\overline{L}_{L}\rho \xi \right) _{\mathbf{\mathbf{1}}_{0}}e_{R}$ operator.
That aforementioned twelve dimensional charged lepton Yukawa term is crucial
to explain the smallness of the electron mass, without tuning its
corresponding Yukawa coupling. 

\quad To get a predictive model that successfully accounts for fermion
masses and mixings, we assume that the $SU(3)_{L}$ singlet scalars have the
following VEV pattern: 
\begin{equation}
\left\langle \sigma \right\rangle =v_{\sigma }e^{i\phi },\hspace{1cm}%
\left\langle \tau \right\rangle =v_{\tau },\hspace{1cm}\left\langle \xi
\right\rangle =\frac{v_{\xi }}{\sqrt{3}}\left( 1,1,1\right) ,\hspace{1cm}%
\left\langle \zeta \right\rangle =\frac{v_{\zeta }}{\sqrt{2}}\left(
1,0,1\right) ,\hspace{1cm}\left\langle S\right\rangle =\frac{v_{S}}{\sqrt{3}}%
\left( 1,1,-1\right) .  \label{VEV}
\end{equation}

We justify this choice of directions in the $T_{7}$ space by the observation
that they describe a natural solution of the scalar potential minimization
equations. Indeed, in the single-field case, $T_{7}$ invariance readily
favors the $(1,1,1)$ direction over e.g. the $(1,0,0)$ solution for large
regions of parameter space. The vacuum $\left\langle \xi \right\rangle $ is
a configuration that preserves a $Z_{3}$ subgroup of $T_{7}$, which has been
extensively studied by many authors, (see for example Ref \cite{T7}). In the
next subsection, we consider the minimization conditions of the high energy
scalar potential (\ref{V1}) of our model, and show that our chosen VEV
directions for the two $T_{7}$ triplets, i.e., $\xi $, $S$ and the $T_{7}$
antitriplet $\zeta $ scalars in Eq. (\ref{VEV}), are consistent with a
global minimum of this scalar potential.

\quad Besides that, the $SU(3)_{L}$ scalar singlets are assumed to acquire
vacuum expectation values at a very high energy $\Lambda _{int}\gg v_{\chi
}\approx \mathcal{O}(1)$ TeV, 
excepting $\zeta _{j}$ ($j=1,2,3$), whose vacuum expectation value is much
lower than the scale of electroweak symmetry breaking $v=246$ GeV. Let's
note that at the scale $\Lambda _{int}$, the $SU(3)_{C}\otimes
SU(3)_{L}\otimes U(1)_{X}\otimes T_{7}\otimes Z_{2}\otimes Z_{3}\otimes
Z_{14}$ symmetry is broken to $SU(3)_{C}\otimes SU(3)_{L}\otimes U(1)_{X}$
by the vacuum expectation values of the $SU(3)_{L}$ singlet scalar fields $%
\xi _{j}$, $S_{j}$, $\sigma $ and $\tau $. 

\quad Considering that the charged fermion mass and quark mixing pattern
arises from the $Z_{2}\otimes Z_{3}\otimes Z_{14}$ symmetry breaking, we set
the VEVs of the $SU(3)_{L}$ singlet scalars $S$, $\xi $, $\sigma $ and $\tau 
$, as follows: 
\begin{equation}
v_{S}\sim v_{\xi }=v_{\sigma }=v_{\tau }=\Lambda _{int}=\lambda \Lambda ,
\label{VEVsinglets}
\end{equation}%
being $\lambda =0.225$ one of the Wolfenstein parameters and $\Lambda $ our
model cutoff. Consequently, the VEVs of the scalars in our model have the
following hierarchy: 
\begin{equation}
v_{\zeta }<<v_{\rho }\sim v_{\eta }\sim v<<v_{\chi }<<\Lambda _{int}.
\end{equation}%
Thus, the $SU(3)_{L}$ scalar singlets having Yukawa interactions with the
right handed Majorana neutrinos get VEVs at very high scale, then providing
very large masses to these Majorana neutrinos, and thus giving rise to a
double seesaw mechanism of active neutrino masses. 
Consequently, the neutrino spectrum includes very light active neutrinos as
well as heavy and very heavy sterile neutrinos. As we will shown in detail
in the next section, the smallness of the active neutrino masses is
attributed to their scaling with inverse powers of the high energy cutoff $%
\Lambda $ as well as by their quadratic dependence on the very small VEV of
the $Z_{2}\otimes Z_{3}\otimes Z_{14}$ neutral, $SU(3)_{L}$ singlet and $%
T_{7}$ antitriplet scalar field $\zeta $.

\bigskip

\subsection{High energy scalar potential}

As previously mentioned, all singlet scalars, excepting $\zeta _{j}$ ($%
j=1,2,3$), acquire vaccum expectation values, much larger than $v_{\chi }$,
which implies that the singlet scalars are very heavy and thus the mixing
between these scalar singlets and the $SU\left( 3\right) _{L}$ scalar
triplets can be neglected as done in Ref. \cite{331S3quarks}. For simplicity
we assume a CP invariant scalar potential with only real couplings as done
in Refs. \cite{M-O,Machado:2010uc,331S3leptons,331S3quarks}. The high energy
scalar potential which involves only the $SU\left( 3\right) _{L}$ singlet
slcalars is given by:

\begin{eqnarray}
V_{1} &=&\mu _{\xi }^{2}\left( \xi \xi ^{\ast }\right) _{\mathbf{\mathbf{1}}%
_{0}}+\mu _{S}^{2}\left( SS^{\ast }\right) _{\mathbf{\mathbf{1}}_{0}}+\mu
_{\zeta }^{2}\left( \zeta \zeta ^{\ast }\right) _{\mathbf{\mathbf{1}}%
_{0}}+\mu _{\sigma }^{2}\left( \sigma \sigma ^{\ast }\right) +\mu _{\tau
}^{2}\left( \tau \tau ^{\ast }\right)  \notag \\
&&+\left[ C_{1}\left( \xi \xi ^{\ast }\right) _{\mathbf{3}}\zeta
+C_{2}\left( SS^{\ast }\right) _{\mathbf{3}}\zeta +C_{3}\left( \xi S\right)
_{\mathbf{3}}\zeta +C_{4}\left( \xi \zeta \right) _{\mathbf{3}}\xi ^{\ast
}+C_{5}\left( S\zeta \right) _{\mathbf{3}}S^{\ast }+C_{6}\left( \xi ^{\ast
}\zeta \right) _{\overline{\mathbf{3}}}\xi \right.  \notag \\
&&+\left. C_{7}\left( S^{\ast }\zeta \right) _{\overline{\mathbf{3}}%
}S+C_{8}\left( \xi \zeta \right) _{\overline{\mathbf{3}}}S+C_{9}\left(
S\zeta \right) _{\overline{\mathbf{3}}}\xi +C_{10}\left( \zeta \zeta \right)
_{\mathbf{3}}\zeta +C_{11}\left( \zeta \zeta \right) _{\overline{\mathbf{3}}%
}\zeta ^{\ast }+C_{10}\left( \zeta \zeta ^{\ast }\right) _{\mathbf{3}}\zeta
\right.  \notag \\
&&+\left. C_{15}\left( \zeta \zeta ^{\ast }\right) _{\overline{\mathbf{3}}%
}\zeta ^{\ast }+C_{12}\left( \xi S\right) _{\overline{\mathbf{3}}}\zeta
^{\ast }+C_{13}\left( \xi ^{\ast }\zeta \right) _{\mathbf{3}}S^{\ast
}+C_{14}\left( S^{\ast }\zeta \right) _{\mathbf{3}}\xi ^{\ast }+H.c\right]
+\kappa _{1}\left( \xi \xi ^{\ast }\right) _{\mathbf{\mathbf{1}}_{0}}\left(
\xi \xi ^{\ast }\right) _{\mathbf{\mathbf{1}}_{0}}  \notag \\
&&+\kappa _{2}\left( \xi \xi ^{\ast }\right) _{\mathbf{1}_{1}}\left( \xi \xi
^{\ast }\right) _{\mathbf{1}_{2}}+\kappa _{3}\left( \xi \xi ^{\ast }\right)
_{\mathbf{3}}\left( \xi \xi ^{\ast }\right) _{\overline{\mathbf{3}}}+\kappa
_{4}\left( SS^{\ast }\right) _{\mathbf{\mathbf{1}}_{0}}\left( SS^{\ast
}\right) _{\mathbf{\mathbf{1}}_{0}}+\kappa _{5}\left( SS^{\ast }\right) _{%
\mathbf{1}_{1}}\left( SS^{\ast }\right) _{\mathbf{1}_{2}}  \notag \\
&&+\kappa _{6}\left( SS^{\ast }\right) _{\mathbf{3}}\left( SS^{\ast }\right)
_{\overline{\mathbf{3}}}+\kappa _{7}\left( \zeta \zeta ^{\ast }\right) _{%
\mathbf{\mathbf{1}}_{0}}\left( \zeta \zeta ^{\ast }\right) _{\mathbf{\mathbf{%
1}}_{0}}+\kappa _{8}\left( \zeta \zeta ^{\ast }\right) _{\mathbf{1}%
_{1}}\left( \zeta \zeta ^{\ast }\right) _{\mathbf{1}_{2}}+\kappa _{9}\left(
\zeta \zeta ^{\ast }\right) _{\mathbf{3}}\left( \zeta \zeta ^{\ast }\right)
_{\overline{\mathbf{3}}}  \notag \\
&&+\left[ \kappa _{10}\left( \zeta \zeta \right) _{\mathbf{3}}\left( \zeta
\zeta \right) _{\overline{\mathbf{3}}}+\gamma _{75}\left( \xi S\right) _{%
\mathbf{3}}\left( \xi S\right) _{\overline{\mathbf{3}}}+\gamma _{1}\left(
\xi \xi \right) _{\mathbf{3}}\left( SS\right) _{\overline{\mathbf{3}}%
}+\gamma _{2}\left( SS\right) _{\mathbf{3}}\left( \xi \xi \right) _{%
\overline{\mathbf{3}}}+H.c\right]  \notag \\
&&+\gamma _{3}\left( \xi \xi ^{\ast }\right) _{\mathbf{\mathbf{1}}%
_{0}}\left( SS^{\ast }\right) _{\mathbf{\mathbf{1}}_{0}}+\gamma _{4}\left(
\xi \xi ^{\ast }\right) _{\mathbf{1}_{1}}\left( SS^{\ast }\right) _{\mathbf{1%
}_{2}}+\gamma _{5}\left( \xi \xi ^{\ast }\right) _{\mathbf{1}_{2}}\left(
SS^{\ast }\right) _{\mathbf{1}_{1}}+\gamma _{6}\left( \xi \xi ^{\ast
}\right) _{\mathbf{3}}\left( SS^{\ast }\right) _{\overline{\mathbf{3}}} 
\notag \\
&&+\gamma _{7}\left( \xi \xi ^{\ast }\right) _{\overline{\mathbf{3}}}\left(
SS^{\ast }\right) _{\mathbf{3}}+\left[ \gamma _{14}\left( \zeta \zeta
\right) _{\mathbf{3}}\left( SS\right) _{\overline{\mathbf{3}}}+\gamma
_{15}\left( \zeta \zeta \right) _{\mathbf{3}}\left( \xi \xi \right) _{%
\overline{\mathbf{3}}}+H.c\right] +\gamma _{8}\left( \xi S^{\ast }\right) _{%
\mathbf{\mathbf{1}}_{0}}\left( S\xi ^{\ast }\right) _{\mathbf{\mathbf{1}}%
_{0}}  \notag \\
&&+\gamma _{9}\left( \xi S^{\ast }\right) _{\mathbf{1}_{1}}\left( S\xi
^{\ast }\right) _{\mathbf{1}_{2}}+\gamma _{11}\left( \xi S^{\ast }\right) _{%
\mathbf{1}_{2}}\left( S\xi ^{\ast }\right) _{\mathbf{1}_{1}}+\gamma
_{12}\left( \xi S\right) _{\mathbf{3}}\left( \xi ^{\ast }S^{\ast }\right) _{%
\overline{\mathbf{3}}}+\gamma _{13}\left( \xi ^{\ast }S^{\ast }\right) _{%
\mathbf{3}}\left( \xi S\right) _{\overline{\mathbf{3}}}  \notag \\
&&+\gamma _{17}\left( \xi S^{\ast }\right) _{\mathbf{3}}\left( \xi ^{\ast
}S\right) _{\overline{\mathbf{3}}}+\gamma _{18}\left( \xi ^{\ast }S\right) _{%
\mathbf{3}}\left( \xi S^{\ast }\right) _{\overline{\mathbf{3}}}+\gamma
_{19}\left( \xi \xi ^{\ast }\right) _{\mathbf{\mathbf{1}}_{0}}\left( \zeta
\zeta ^{\ast }\right) _{\mathbf{\mathbf{1}}_{0}}+\gamma _{20}\left( \xi \xi
^{\ast }\right) _{\mathbf{1}_{1}}\left( \zeta \zeta ^{\ast }\right) _{%
\mathbf{1}_{2}}  \notag \\
&&+\gamma _{21}\left( \xi \xi ^{\ast }\right) _{\mathbf{1}_{2}}\left( \zeta
\zeta ^{\ast }\right) _{\mathbf{1}_{1}}+\gamma _{23}\left( \xi \xi ^{\ast
}\right) _{\mathbf{3}}\left( \zeta \zeta ^{\ast }\right) _{\overline{\mathbf{%
3}}}+\gamma _{24}\left( \zeta \zeta ^{\ast }\right) _{\mathbf{3}}\left( \xi
\xi ^{\ast }\right) _{\overline{\mathbf{3}}}  \notag \\
&&+\left[ \gamma _{25}\left( \xi \xi ^{\ast }\right) _{\mathbf{3}}\left(
\zeta \zeta \right) _{\overline{\mathbf{3}}}+\gamma _{26}\left( \zeta \zeta
\right) _{\mathbf{3}}\left( \xi \xi ^{\ast }\right) _{\overline{\mathbf{3}}%
}+H.c\right] +\gamma _{30}\left( \xi \zeta \right) _{\mathbf{3}}\left( \xi
^{\ast }\zeta ^{\ast }\right) _{\overline{\mathbf{3}}}  \notag \\
&&+\gamma _{27}\left( \xi \zeta \right) _{\mathbf{\mathbf{1}}_{0}}\left( \xi
^{\ast }\zeta ^{\ast }\right) _{\mathbf{\mathbf{1}}_{0}}+\gamma _{28}\left(
\xi \zeta \right) _{\mathbf{1}_{1}}\left( \xi ^{\ast }\zeta ^{\ast }\right)
_{\mathbf{1}_{2}}+\gamma _{29}\left( \xi \zeta \right) _{\mathbf{1}%
_{2}}\left( \xi ^{\ast }\zeta ^{\ast }\right) _{\mathbf{1}_{1}}  \notag \\
&&+\gamma _{31}\left( \xi ^{\ast }\zeta ^{\ast }\right) _{\mathbf{3}}\left(
\xi \zeta \right) _{\overline{\mathbf{3}}}+\left[ \gamma _{32}\left( \xi
\zeta \right) _{\mathbf{3}}\left( \xi ^{\ast }\zeta \right) _{\overline{%
\mathbf{3}}}+\gamma _{33}\left( \xi ^{\ast }\zeta \right) _{\mathbf{3}%
}\left( \xi \zeta \right) _{\overline{\mathbf{3}}}+H.c\right]  \notag \\
&&+\gamma _{34}\left( SS^{\ast }\right) _{\mathbf{\mathbf{1}}_{0}}\left(
\zeta \zeta ^{\ast }\right) _{\mathbf{\mathbf{1}}_{0}}+\gamma _{35}\left(
SS^{\ast }\right) _{\mathbf{1}_{1}}\left( \zeta \zeta ^{\ast }\right) _{%
\mathbf{1}_{2}}+\gamma _{36}\left( SS^{\ast }\right) _{\mathbf{1}_{2}}\left(
\zeta \zeta ^{\ast }\right) _{\mathbf{1}_{1}}+\gamma _{37}\left( SS^{\ast
}\right) _{\mathbf{3}}\left( \zeta \zeta ^{\ast }\right) _{\overline{\mathbf{%
3}}}  \notag \\
&&+\gamma _{38}\left( \zeta \zeta ^{\ast }\right) _{\mathbf{3}}\left(
SS^{\ast }\right) _{\overline{\mathbf{3}}}+\left[ \gamma _{39}\left(
SS^{\ast }\right) _{\mathbf{3}}\left( \zeta \zeta \right) _{\overline{%
\mathbf{3}}}+\gamma _{40}\left( \zeta \zeta \right) _{\mathbf{3}}\left(
SS^{\ast }\right) _{\overline{\mathbf{3}}}+H.c\right] +\gamma _{44}\left(
S\zeta \right) _{\mathbf{3}}\left( S^{\ast }\zeta ^{\ast }\right) _{%
\overline{\mathbf{3}}}  \notag \\
&&+\gamma _{41}\left( S\zeta \right) _{\mathbf{\mathbf{1}}_{0}}\left(
S^{\ast }\zeta ^{\ast }\right) _{\mathbf{\mathbf{1}}_{0}}+\gamma _{42}\left(
S\zeta \right) _{\mathbf{1}_{1}}\left( S^{\ast }\zeta ^{\ast }\right) _{%
\mathbf{1}_{2}}+\gamma _{43}\left( S\zeta \right) _{\mathbf{1}_{2}}\left(
S^{\ast }\zeta ^{\ast }\right) _{\mathbf{1}_{1}}+\gamma _{45}\left( S^{\ast
}\zeta ^{\ast }\right) _{\mathbf{3}}\left( S\zeta \right) _{\overline{%
\mathbf{3}}}  \notag \\
&&+\left[ \gamma _{49}\left( \xi S\right) _{\mathbf{3}}\left( \zeta \zeta
^{\ast }\right) _{\overline{\mathbf{3}}}+\gamma _{50}\left( \zeta \zeta
^{\ast }\right) _{\mathbf{3}}\left( \xi S\right) _{\overline{\mathbf{3}}%
}+\gamma _{51}\left( \xi S\right) _{\mathbf{3}}\left( \zeta \zeta \right) _{%
\overline{\mathbf{3}}}+\gamma _{52}\left( \zeta \zeta \right) _{\mathbf{3}%
}\left( \xi S\right) _{\overline{\mathbf{3}}}+H.c\right]  \notag \\
&&+\left[ \gamma _{53}\left( \xi \zeta \right) _{\mathbf{\mathbf{1}}%
_{0}}\left( S\zeta \right) _{\mathbf{\mathbf{1}}_{0}}+\gamma _{54}\left( \xi
\zeta \right) _{\mathbf{1}_{1}}\left( S\zeta \right) _{\mathbf{1}%
_{2}}+\gamma _{55}\left( \xi \zeta \right) _{\mathbf{1}_{2}}\left( S\zeta
\right) _{\mathbf{1}_{1}}\right.  \notag \\
&&+\left. \gamma _{56}\left( \xi \zeta \right) _{\mathbf{3}}\left( S\zeta
\right) _{\overline{\mathbf{3}}}+\gamma _{57}\left( S\zeta \right) _{\mathbf{%
3}}\left( \xi \zeta \right) _{\overline{\mathbf{3}}}+H.c\right] +\gamma
_{66}\left( \zeta \zeta ^{\ast }\right) _{\mathbf{\mathbf{1}}_{0}}\left(
\sigma \sigma ^{\ast }\right) +\gamma _{16}\left( \zeta \zeta ^{\ast
}\right) _{\mathbf{\mathbf{1}}_{0}}\left( \sigma \sigma ^{\ast }\right) 
\notag \\
&&+\gamma _{58}\left( \xi \xi ^{\ast }\right) _{\mathbf{\mathbf{1}}%
_{0}}\left( \sigma \sigma ^{\ast }\right) +\gamma _{59}\left( SS^{\ast
}\right) _{\mathbf{\mathbf{1}}_{0}}\left( \sigma \sigma ^{\ast }\right)
+\gamma _{16}\left( \zeta \zeta ^{\ast }\right) _{\mathbf{1}_{1}}\zeta
^{2}+\gamma _{67}\left( \zeta \zeta ^{\ast }\right) _{\mathbf{1}_{2}}\zeta
^{\ast 2}  \notag \\
&&+\gamma _{61}\left( \xi \xi ^{\ast }\right) _{\mathbf{\mathbf{1}}%
_{0}}\left( \tau \tau ^{\ast }\right) +\gamma _{62}\left( SS^{\ast }\right)
_{\mathbf{\mathbf{1}}_{0}}\left( \tau \tau ^{\ast }\right) +\gamma
_{68}\left( \sigma \sigma ^{\ast }\right) ^{2}+\gamma _{69}\left( \tau \tau
^{\ast }\right) ^{2}+\gamma _{70}\left( \sigma \sigma ^{\ast }\right) \left(
\tau \tau ^{\ast }\right)  \notag \\
&&+\gamma _{71}\left( \xi \xi ^{\ast }\right) _{\mathbf{1}_{1}}\tau
^{2}+\gamma _{72}\left( \xi \xi ^{\ast }\right) _{\mathbf{1}_{2}}\tau ^{\ast
2}+\gamma _{73}\left( SS^{\ast }\right) _{\mathbf{1}_{1}}\tau ^{2}+\gamma
_{74}\left( SS^{\ast }\right) _{\mathbf{1}_{2}}\tau ^{\ast 2}  \label{V1}
\end{eqnarray}%
Now we are going to determine the conditions under which the VEV pattern for
the $SU(3)_{L}$ singlet scalars, given in Eq. (\ref{VEV}), is a solution of
the high energy scalar potential. In view of the very large number of
parameters of the high energy scalar potential, in order to simplify the
analysis, we assume universality in its trilinear and quartic couplings,
i.e. 
\begin{eqnarray}
\kappa _{i} &=&\gamma _{j}=\kappa ,\ \ \ \ \ \ \ \ \ i=(1-10),\ \ \ \ \ \ \
\ \ j=(1-75),  \notag \\
C_{k} &=&C,\ \ \ \ \ \ \ \ \ k=(1-15)
\end{eqnarray}%
Then, from the minimization conditions of the scalar potential and taking
into account our assumption that $v_{\zeta _{j}}$ ($j=1,2,3$) are much lower
than the electroweak symmetry breaking scale $v=246$ GeV, the following
relations are obtained: 
\begin{eqnarray}
\frac{\partial \left\langle V_{1}\right\rangle }{\partial v_{\xi _{1}}}
&=&8\kappa v_{\xi _{1}}^{3}+4\kappa v_{\xi _{2}}^{3}+2v_{\xi _{1}}\mu _{\xi
}^{2}+2\kappa v_{\xi _{1}}\left( v_{\xi _{2}}^{2}+v_{\xi _{3}}^{2}\right)
+12\kappa v_{\xi _{1}}^{2}v_{\xi _{3}}+2\kappa v_{\xi _{1}}v_{\xi
_{2}}^{2}+2\kappa v_{S_{1}}^{2}\left( 7v_{\xi _{1}}+3v_{\xi _{3}}\right) 
\notag \\
&&+2\kappa v_{S_{2}}^{2}\left( 2v_{\xi _{1}}+3v_{\xi _{3}}\right) +4\kappa
v_{S_{3}}^{2}v_{\xi _{1}}+2\kappa \left( v_{\sigma }^{2}+3v_{\tau
}^{2}\right) v_{\xi _{1}}+4\kappa v_{S_{1}}\left( v_{\xi
_{2}}v_{S_{2}}+v_{\xi _{3}}v_{S_{3}}+3v_{\xi _{1}}v_{S_{3}}\right)  \notag \\
&=&0,  \notag \\
\frac{\partial \left\langle V_{1}\right\rangle }{\partial v_{\xi _{2}}}
&=&8\kappa v_{\xi _{2}}^{3}+4\kappa v_{\xi _{3}}^{3}+2v_{\xi _{2}}\mu _{\xi
}^{2}+2\kappa v_{\xi _{2}}\left( v_{\xi _{1}}^{2}+v_{\xi _{3}}^{2}\right)
+12\kappa v_{\xi _{2}}^{2}v_{\xi _{1}}+2\kappa v_{\xi _{2}}v_{\xi
_{3}}^{2}+2\kappa v_{S_{2}}^{2}\left( 7v_{\xi _{2}}+3v_{\xi _{1}}\right) 
\notag \\
&&+2\kappa v_{S_{3}}^{2}\left( 2v_{\xi _{2}}+3v_{\xi _{3}}\right) +4\kappa
v_{S_{1}}^{2}v_{\xi _{2}}+2\kappa v_{\sigma }^{2}v_{\xi _{2}}+4\kappa
v_{S_{2}}\left( v_{\xi _{1}}v_{S_{1}}+v_{\xi _{3}}v_{S_{3}}+3v_{\xi
_{2}}v_{S_{1}}\right)  \notag \\
&=&0,  \notag \\
\frac{\partial \left\langle V_{1}\right\rangle }{\partial v_{\xi _{3}}}
&=&8\kappa v_{\xi _{3}}^{3}+4\kappa v_{\xi _{1}}^{3}+2v_{\xi _{3}}\mu _{\xi
}^{2}+2\kappa v_{\xi _{3}}\left( v_{\xi _{1}}^{2}+v_{\xi _{2}}^{2}\right)
+12\kappa v_{\xi _{3}}^{2}v_{\xi _{2}}+2\kappa v_{\xi _{3}}v_{\xi
_{2}}^{2}+2\kappa v_{S_{3}}^{2}\left( 7v_{\xi _{3}}+3v_{\xi _{2}}\right) 
\notag \\
&&+2\kappa v_{S_{1}}^{2}\left( 2v_{\xi _{3}}+3v_{\xi _{1}}\right) +4\kappa
v_{S_{2}}^{2}v_{\xi _{3}}+2\kappa v_{\sigma }^{2}v_{\xi _{3}}+4\kappa
v_{S_{3}}\left( v_{\xi _{2}}v_{S_{2}}+v_{\xi _{1}}v_{S_{1}}+3v_{\xi
_{3}}v_{S_{2}}\right)  \notag \\
&=&0,  \label{DVxi}
\end{eqnarray}

\begin{eqnarray}
\frac{\partial \left\langle V_{1}\right\rangle }{\partial \text{$v_{S_{1}}$}}
&=&8\kappa v_{S_{1}}^{3}+4\kappa v_{S_{2}}^{3}+2v_{S_{1}}\mu _{\xi
}^{2}+2\kappa v_{S_{1}}\left( v_{S_{2}}^{2}+v_{S_{3}}^{2}\right) +12\kappa
v_{S_{1}}^{2}v_{S_{3}}+2\kappa v_{S_{1}}v_{S_{2}}^{2}+2\kappa v_{\xi
_{1}}^{2}\left( 7v_{S_{1}}+3v_{S_{3}}\right)  \notag \\
&&+2\kappa v_{\xi _{2}}^{2}\left( 2v_{S_{1}}+3v_{S_{3}}\right) +4\kappa
v_{\xi _{3}}^{2}v_{S_{1}}+2\kappa \left( v_{\sigma }^{2}+3v_{\tau
}^{2}\right) v_{S_{1}}+4\kappa v_{\xi _{1}}\left( v_{S_{2}}v_{\xi
_{2}}+v_{S_{3}}v_{\xi _{3}}+3v_{S_{1}}v_{\xi _{3}}\right)  \notag \\
&=&0,  \notag \\
\frac{\partial \left\langle V_{1}\right\rangle }{\partial \text{$v_{S_{2}}$}}
&=&8\kappa v_{S_{2}}^{3}+4\kappa v_{S_{3}}^{3}+2v_{S_{2}}\mu _{\xi
}^{2}+2\kappa v_{S_{2}}\left( v_{S_{1}}^{2}+v_{S_{3}}^{2}\right) +12\kappa
v_{S_{2}}^{2}v_{S_{1}}+2\kappa v_{S_{2}}v_{S_{3}}^{2}+2\kappa v_{\xi
_{2}}^{2}\left( 7v_{S_{2}}+3v_{S_{1}}\right)  \notag \\
&&+2\kappa v_{\xi _{3}}^{2}\left( 2v_{S_{2}}+3v_{S_{3}}\right) +4\kappa
v_{\xi _{1}}^{2}v_{S_{2}}+2\kappa v_{\sigma }^{2}v_{S_{2}}+4\kappa v_{\xi
_{2}}\left( v_{S_{1}}v_{\xi _{1}}+v_{S_{3}}v_{\xi _{3}}+3v_{S_{2}}v_{\xi
_{1}}\right)  \notag \\
&=&0,  \notag \\
\frac{\partial \left\langle V_{1}\right\rangle }{\partial \text{$v_{S_{3}}$}}
&=&8\kappa v_{S_{3}}^{3}+4\kappa v_{S_{1}}^{3}+2v_{S_{3}}\mu _{\xi
}^{2}+2\kappa v_{S_{3}}\left( v_{S_{1}}^{2}+v_{S_{2}}^{2}\right) +12\kappa
v_{S_{3}}^{2}v_{S_{2}}+2\kappa v_{S_{3}}v_{S_{2}}^{2}+2\kappa v_{\xi
_{3}}^{2}\left( 7v_{S_{3}}+3v_{S_{2}}\right)  \notag \\
&&+2\kappa v_{\xi _{1}}^{2}\left( 2v_{S_{3}}+3v_{S_{1}}\right) +4\kappa
v_{\xi _{2}}^{2}v_{S_{3}}+2\kappa v_{\sigma }^{2}v_{S_{3}}+4\kappa v_{\xi
_{3}}\left( v_{S_{2}}v_{\xi _{2}}+v_{S_{1}}v_{\xi _{1}}+3v_{S_{3}}v_{\xi
_{2}}\right)  \notag \\
&=&0,
\end{eqnarray}

\begin{eqnarray}
\frac{\partial \left\langle V_{1}\right\rangle }{\partial \text{$v_{\zeta
_{1}}$}} &=&6C\left[ v_{\xi _{1}}v_{\xi _{2}}+\text{$v_{S_{1}}$}%
v_{S_{2}}+2\left( v_{\xi _{3}}v_{S_{2}}+v_{\xi _{2}}v_{S_{3}}\right) +v_{\xi
_{3}}v_{S_{3}}\right] +2\text{$\mu _{\zeta }^{2}v_{\zeta _{1}}$}=0,  \notag
\\
\frac{\partial \left\langle V_{1}\right\rangle }{\partial \text{$v_{\zeta
_{2}}$}} &=&6C\left[ v_{\xi _{2}}v_{\xi _{3}}+\text{$v_{S_{2}}$}v_{S_{3}}%
\text{$+2\left( v_{\xi _{3}}v_{S_{1}}+v_{\xi _{1}}v_{S_{3}}\right) +$}v_{\xi
_{1}}v_{S_{1}}\right] +2\text{$\mu _{\zeta }^{2}v_{\zeta _{2}}=0,$}  \notag
\\
\frac{\partial \left\langle V_{1}\right\rangle }{\partial \text{$v_{\zeta
_{3}}$}} &=&6C\left[ v_{\xi _{1}}v_{\xi _{3}}+v_{S_{1}}\text{$%
v_{S_{3}}+2\left( v_{\xi _{2}}v_{S_{1}}+v_{\xi _{1}}v_{S_{2}}\right) +$}%
v_{\xi _{2}}v_{S_{2}}\right] +2\text{$\mu _{\zeta }^{2}v_{\zeta _{3}}=0,$}
\end{eqnarray}%
\begin{eqnarray}
&&\frac{\partial \left\langle V_{1}\right\rangle }{\partial \text{$v_{\sigma
}$}}=2v_{\sigma }\left[ \mu _{\sigma }^{2}+\kappa \left( v_{\xi
_{1}}^{2}+v_{\xi _{2}}^{2}+v_{\xi
_{3}}^{2}+v_{S_{1}}^{2}+v_{S_{2}}^{2}+v_{S_{3}}^{2}+2v_{\sigma }^{2}+v_{\tau
}^{2}\right) \right] =0,  \notag \\
&&\frac{\partial \left\langle V_{1}\right\rangle }{\partial \text{$v_{\tau }$%
}}=2v_{\tau }\left[ \mu _{\tau }^{2}+\kappa \left( 3v_{\xi
_{1}}^{2}+3v_{S_{1}}^{2}+v_{\sigma }^{2}+2v_{\tau }^{2}\right) \right] =0.
\end{eqnarray}%
From the expressions given above, and using the vacuum configuration for the 
$SU(3)_{L}$ singlet scalars given in Eq. (\ref{VEV}), we find the following
relations: 
\begin{eqnarray}
\mu _{\xi }^{2} &=&-\frac{1}{9}\kappa \left( 60v_{\xi
}^{2}+41v_{S}^{2}\right) ,\hspace{1cm}\mu _{S}^{2}=-\frac{1}{9}\kappa \left(
59v_{\xi }^{2}+10v_{S}^{2}\right) ,  \notag \\
\mu _{\zeta }^{2} &=&-\frac{\sqrt{2}Cv_{\xi }}{\text{$v_{\zeta }$}}\left(
v_{\xi }+v_{S}\right) ,\hspace{1cm}\mu _{\sigma }^{2}=\mu _{\tau
}^{2}=-\kappa \left( 4v_{\xi }^{2}+v_{S}^{2}\right) ,  \notag \\
&&2v_{\xi }^{2}-2v_{S}^{2}+v_{\xi }v_{S}=0.
\end{eqnarray}%
Taking the positive solution of the previous equation, we find: 
\begin{equation}
v_{S}=\frac{1+\sqrt{5}}{4}v_{\xi }\simeq 0.81v_{\xi },
\end{equation}%
which is consistent with our previous assumption described by Eq. (\ref%
{VEVsinglets}). Our results show that the VEV directions for the two $T_{7}$
triplets, i.e., $\xi $, $S$ and the $T_{7}$ antitriplet $\zeta $ scalars in
Eq. (\ref{VEV}), are consistent with a global minimum of the high scalar
potential (\ref{V1}) of our model, for a not fine-tuned region of parameter
space.

\subsection{Low energy scalar potential}

The renormalizable low energy scalar potential of the model takes the form:%
%
%
%
%
%
%
%
%
%
%
\begin{eqnarray}
&&V_{H}=\mu _{\chi }^{2}(\chi ^{\dagger }\chi )+\mu _{\eta }^{2}(\eta
^{\dagger }\eta )+\mu _{\rho }^{2}(\rho ^{\dagger }\rho )+f\left( \chi
_{i}\eta _{j}\rho _{k}\varepsilon ^{ijk}+H.c.\right) +\lambda _{1}(\chi
^{\dagger }\chi )(\chi ^{\dagger }\chi )  \notag \\
&&+\lambda _{2}(\rho ^{\dagger }\rho )(\rho ^{\dagger }\rho )+\lambda
_{3}(\eta ^{\dagger }\eta )(\eta ^{\dagger }\eta )+\lambda _{4}(\chi
^{\dagger }\chi )(\rho ^{\dagger }\rho )+\lambda _{5}(\chi ^{\dagger }\chi
)(\eta ^{\dagger }\eta )  \notag \\
&&+\lambda _{6}(\rho ^{\dagger }\rho )(\eta ^{\dagger }\eta )+\lambda
_{7}(\chi ^{\dagger }\eta )(\eta ^{\dagger }\chi )+\lambda _{8}(\chi
^{\dagger }\rho )(\rho ^{\dagger }\chi )+\lambda _{9}(\rho ^{\dagger }\eta
)(\eta ^{\dagger }\rho ).  \label{v00}
\end{eqnarray}

After the symmetry breaking, it is found that the scalar mass eigenstates
are connected with the weak scalar states by the following relations: \cite%
{331-long, M-O}: 
\begin{eqnarray}
\begin{pmatrix}
G_{1}^{\pm } \\ 
H_{1}^{\pm } \\ 
\end{pmatrix}%
=R_{\beta _{T}}%
\begin{pmatrix}
\rho _{1}^{\pm } \\ 
\eta _{2}^{\pm } \\ 
\end{pmatrix}
&,&\hspace{0.3cm}%
\begin{pmatrix}
G_{1}^{0} \\ 
A_{1}^{0} \\ 
\end{pmatrix}%
=R_{\beta _{T}}%
\begin{pmatrix}
\zeta _{\rho } \\ 
\zeta _{\eta } \\ 
\end{pmatrix}%
,\hspace{0.3cm}%
\begin{pmatrix}
H_{1}^{0} \\ 
h^{0} \\ 
\end{pmatrix}%
=R_{\alpha _{T}}%
\begin{pmatrix}
\xi _{\rho } \\ 
\xi _{\eta } \\ 
\end{pmatrix}%
,  \label{331-mass-scalar-a} \\
\begin{pmatrix}
G_{2}^{0} \\ 
H_{2}^{0} \\ 
\end{pmatrix}%
=R%
\begin{pmatrix}
\chi _{1}^{0} \\ 
\eta _{3}^{0} \\ 
\end{pmatrix}
&,&\hspace{0.3cm}%
\begin{pmatrix}
G_{3}^{0} \\ 
H_{3}^{0} \\ 
\end{pmatrix}%
=R%
\begin{pmatrix}
\zeta _{\chi } \\ 
\xi _{\chi } \\ 
\end{pmatrix}%
,\hspace{0.3cm}%
\begin{pmatrix}
G_{2}^{\pm } \\ 
H_{2}^{\pm } \\ 
\end{pmatrix}%
=R%
\begin{pmatrix}
\chi _{2}^{\pm } \\ 
\rho _{3}^{\pm } \\ 
\end{pmatrix}%
,  \label{331-mass-scalar-b}
\end{eqnarray}

with

\begin{equation}
R_{\alpha _{T}(\beta _{T})}=\left( 
\begin{array}{cc}
\cos \alpha _{T}(\beta _{T}) & \sin \alpha _{T}(\beta _{T}) \\ 
-\sin \alpha _{T}(\beta _{T}) & \cos \alpha _{T}(\beta _{T})%
\end{array}%
\right) ,\hspace{2cm}R=\left( 
\begin{array}{cc}
-1 & 0 \\ 
0 & 1%
\end{array}%
\right) ,
\end{equation}
where $\tan \beta _{T}=v_{\eta }/v_{\rho }$, and $\tan 2\alpha
_{T}=M_1/(M_2-M_3)$ with: 
\begin{eqnarray}
M_1&=&4\lambda _6 v_{\eta}v_{\rho}+2\sqrt{2}fv_{\chi},  \notag \\
M_2&=&4\lambda _2 v_{\rho}^2-\sqrt{2}fv_{\chi}\tan \beta _T ,  \notag \\
M_3&=&4\lambda _3 v_{\eta}^2-\sqrt{2}fv_{\chi}/\tan \beta _T .
\end{eqnarray}

The low energy physical scalar spectrum of our model includes: 4 massive
charged Higgs ($H_{1}^{\pm }$, $H_{2}^{\pm }$), one CP-odd Higgs ($A_{1}^{0}$%
), 3 neutral CP-even Higgs ($h^{0},H_{1}^{0},H_{3}^{0}$) and 2 neutral Higgs
($H_{2}^{0},\overline{H}_{2}^{0}$) bosons. The scalar $h^{0}$ is identified
with the SM-like $126$ GeV Higgs boson found at the LHC. It it noteworthy
that the neutral Goldstone bosons $G_{1}^{0}$, $G_{3}^{0}$, $G_{2}^{0}$ , $%
\overline{G}_{2}^{0}$ are associated to the longitudinal components of the $%
Z $, $Z^{\prime }$, $K^{0}$ and $\overline{K}^{0}$gauge bosons,
respectively. Furthermore, the charged Goldstone bosons $G_{1}^{\pm }$ and $%
G_{2}^{\pm }$ are associated to the longitudinal components of the $W^{\pm }$
and $K^{\pm } $ gauge bosons, respectively \cite{331-pisano,M-O}.

%
%

\section{Lepton masses and mixings}

\label{leptonmassesandmixing} From Eqs. (\ref{Lylepton}), (\ref{VEV}), (\ref%
{VEVsinglets}) and using the product rules of the $T_{7}$ group given in
Appendix \ref{ap1}, it follows that the mass matrix for charged leptons is: 
\begin{equation}
M_{l}=V_{lL}^{\dag }P_{l}diag\left( m_{e},m_{\mu },m_{\tau }\right) ,\hspace{%
0.5cm}V_{lL}=\frac{1}{\sqrt{3}}\left( 
\begin{array}{ccc}
1 & 1 & 1 \\ 
1 & \omega & \omega ^{2} \\ 
1 & \omega ^{2} & \omega%
\end{array}%
\right) ,\hspace{0.5cm}P_{l}=\left( 
\begin{array}{ccc}
e^{7i\phi } & 0 & 0 \\ 
0 & e^{4i\phi } & 0 \\ 
0 & 0 & e^{2i\phi }%
\end{array}%
\right) ,\hspace{0.5cm}\omega =e^{\frac{2\pi i}{3}},
\end{equation}%
%
%
%
%
%
%
%
%
%
%
%
%
%
%
%
%
%
%
%
%
%
%
%
%
%
%
%
%
%
%
where the charged lepton masses read:
\begin{equation}
m_{e}=h_{\rho e}^{\left( L\right) }\lambda ^{8}\frac{v_{\rho }}{\sqrt{2}},%
\hspace{1cm}m_{\mu }=h_{\rho \mu }^{\left( L\right) }\lambda ^{5}\frac{%
v_{\rho }}{\sqrt{2}},\hspace{1cm}m_{\tau }=h_{\rho \tau }^{\left( L\right)
}\lambda ^{3}\frac{v_{\rho }}{\sqrt{2}}.  \label{leptonmasses}
\end{equation}%
Taking into account that $v_{\rho }\approx v=246$ GeV, it follows that the
charged lepton masses are related with the electroweak symmetry breaking
scale by their scalings with powers of the Wolfenstein parameter $\lambda
=0.225$, with $\mathcal{O}(1)$ coefficients. 

\quad In the concerning to the neutrino sector, the following neutrino mass
terms arise: 
\begin{equation}
-\mathcal{L}_{mass}^{\left( \nu \right) }=\frac{1}{2}\left( 
\begin{array}{ccc}
\overline{\nu _{L}^{C}} & \overline{\nu _{R}} & \overline{N_{R}}%
\end{array}%
\right) M_{\nu }\left( 
\begin{array}{c}
\nu _{L} \\ 
\nu _{R}^{C} \\ 
N_{R}^{C}%
\end{array}%
\right) +H.c,  \label{Lnu}
\end{equation}%
where the $T_{7}$ discrete flavor group constrains the neutrino mass matrix
to be of the form: 
\begin{eqnarray}
M_{\nu } &=&\left( 
\begin{array}{ccc}
0_{3\times 3} & M_{D} & 0_{3\times 3} \\ 
M_{D}^{T} & 0_{3\times 3} & M_{\chi } \\ 
0_{3\times 3} & M_{\chi }^{T} & M_{R}%
\end{array}%
\right) ,\hspace{0.7cm}M_{D}=\frac{h_{\rho }v_{\rho }v_{\zeta }}{2\Lambda }%
\left( 
\begin{array}{ccc}
0 & 1 & 0 \\ 
-1 & 0 & -1 \\ 
0 & 1 & 0%
\end{array}%
\right) ,\hspace{0.7cm}M_{\chi }=h_{\chi }^{\left( L\right) }\frac{v_{\chi }%
}{\sqrt{2}}\left( 
\begin{array}{ccc}
1 & 0 & 0 \\ 
0 & 1 & 0 \\ 
0 & 0 & 1%
\end{array}%
\right) ,  \notag \\
M_{R} &=&h_{1N}\frac{v_{\xi }}{\sqrt{3}}\left( 
\begin{array}{ccc}
1 & -x & x \\ 
-x & 1 & x \\ 
x & x & 1%
\end{array}%
\right) ,\hspace{0.7cm}x=\frac{h_{2N}v_{S}}{h_{1N}v_{\xi }}.
\end{eqnarray}

Since the $SU(3)_{L}$ singlet scalars having Yukawa interactions with the
right handed Majorana neutrinos acquire VEVs at very high scale, these
Majorana neutrinos are very heavy, so that the active neutrinos get small
masses via a double seesaw mechanism.

\quad The full rotation matrix, which diagonalizes the neutrino mass matrix,
takes the approximate form \cite{catano}: 
\begin{equation}
\mathbb{U}=%
\begin{pmatrix}
V_{\nu } & B_{2}U_{\chi } & 0 \\ 
-B_{2}^{\dagger }V_{\nu } & U_{\chi } & B_{1}U_{R} \\ 
0 & B_{1}^{\dagger }U_{\chi } & U_{R}%
\end{pmatrix}%
,  \label{U}
\end{equation}%
where 
\begin{equation}
B_{1}^{\dagger }=M_{R}^{-1}M_{\chi }^{T},\hspace{1cm}\hspace{1cm}%
B_{2}^{\dagger }=M_{D}\left( M_{\chi }^{T}\right) ^{-1}M_{R}M_{\chi }^{-1},
\label{B}
\end{equation}%
and the neutrino mass matrices for the physical states are: 
\begin{eqnarray}
M_{\nu }^{\left( 1\right) } &=&M_{D}\left( M_{\chi }^{T}\right)
^{-1}M_{R}M_{\chi }^{-1}M_{D}^{T},  \label{Mnu1} \\
M_{\nu }^{\left( 2\right) } &=&-M_{\chi }M_{R}^{-1}M_{\chi }^{T},\hspace{1cm}%
\hspace{1cm}  \label{Mnu2} \\
M_{\nu }^{\left( 3\right) } &=&M_{R},  \label{Mnu3}
\end{eqnarray}
being $M_{\nu }^{\left( 1\right) }$ the mass matrix for light active
neutrinos, while $M_{\nu }^{\left( 2\right) }$ and $M_{\nu }^{\left(
3\right) }$ are the heavy and very heavy sterile neutrino mass matrices,
respectively. Consequently, the double seesaw mechanism gives rise to light
active neutrinos as well as to heavy and very heavy sterile neutrinos.
Moreover, the neutrino mass matrices $M_{\nu }^{\left( 1\right) }$, $M_{\nu
}^{\left( 2\right) }$ and $M_{\nu }^{\left( 3\right) }$ are diagonalized by
the rotation matrices $V_{\nu }$, $U_{R}$ and $U_{\chi }$, respectively. 
\cite{catano}.

\quad Using Eq. (\ref{Mnu1}), we find the following mass matrix for light
active neutrinos: 
\begin{equation}
M_{\nu }^{\left( 1\right) }=\left( 
\begin{array}{ccc}
A & 0 & A \\ 
0 & B & 0 \\ 
A & 0 & A%
\end{array}%
\right) ,  \label{Mnu}
\end{equation}%
where 
\begin{equation}
A=\frac{h_{1N}h_{\rho }^{2}v_{\rho }^{2}v_{\zeta }^{2}v_{\xi }}{2\sqrt{3}%
h_{\chi }^{\left( L\right) }v_{\chi }^{2}\Lambda ^{2}},\hspace{1cm}\hspace{%
1cm}B=\frac{h_{\rho }^{2}v_{\rho }^{2}v_{\zeta }^{2}}{\sqrt{3}h_{\chi
}^{\left( L\right) }v_{\chi }^{2}\Lambda ^{2}}\left( h_{1N}v_{\xi
}+h_{2N}v_{S}\right) .
\end{equation}%
From Eq. (\ref{Mnu}) it follows that the light active neutrino mass matrix 
only depends on two effective parameters: $A$ and $B$, which determine the
neutrino mass squared splittings. 
Let's note that $A$ and $B$ are supressed by their scaling with inverse
powers of the high energy cutoff $\Lambda $. Furthermore, we have that the
smallness of the active neutrino masses arises from their scaling with
inverse powers of the high energy cutoff $\Lambda $ as well as from their
quadratic dependence on the very small VEV of the $Z_{2}\otimes Z_{3}\otimes
Z_{14}$ neutral, $SU(3)_{L}$ singlet and $T_{7}$ antitriplet scalar field $%
\zeta $. Considering that the orders of magnitude of the SM particles and
new physics yield the constraints $v_{\chi }\gtrsim 1$ TeV and $v_{\eta
}^{2}+v_{\rho }^{2}=v^{2}$ and taking into account our assumption that the
dimensionless lepton Yukawa couplings are $\mathcal{O}(1)$ parameters, from
Eq. (\ref{Mnu}) and the relations $v_{\xi }=\lambda \Lambda $, $v_{\rho
}\sim 100$ GeV, $v_{\chi }\sim 1$ TeV,\ we get that the mass scale for the
light active neutrinos satisfies $m_{\nu }\sim 10^{-3}\frac{v_{\zeta }^{2}}{%
\Lambda }$. Consequently, setting $v_{\zeta }=1$ GeV, we find for the cutoff
of our model the estimate 
\begin{equation}
\Lambda \sim 10^{5}\text{ TeV},  \label{cutoff}
\end{equation}%
which is of the same order of magnitude of the cutoff of our $S_{3}$ lepton
flavor $331$ model \cite{331S3leptons}. Consequenty, we find that the heavy
and very heavy sterile neutrinos have masses at the $\sim $ MeV and $\sim $
TeV scales, respectively. Furthermore, from the aforementioned
considerations, 
as well as from Eqs (\ref{VEVsinglets}), (\ref{B}) and (\ref{cutoff}), it
follows that: 
\begin{equation}
\left\vert \left( B_{2}\right) _{ij}\right\vert \sim \frac{v_{\rho }v_{\zeta
}v_{\xi }}{v_{\chi }^{2}\Lambda }\sim 10^{-5}<<\left\vert \left(
B_{1}\right) _{ij}\right\vert \sim \frac{v_{\chi }}{v_{\xi }}\sim 10^{-3},%
\hspace{0.5cm}i,j=1,2,3,  \label{Binequality}
\end{equation}%
\quad Moreover, we find that the mass matrix $M_{\nu }^{\left( 1\right) }$
for light active neutrinos is diagonalized by a rotation matrix $V_{\nu }$,
as follows: 
\begin{equation}
V_{\nu }^{T}M_{\nu }^{\left( 1\right) }V_{\nu }=\left( 
\begin{array}{ccc}
m_{1} & 0 & 0 \\ 
0 & m_{2} & 0 \\ 
0 & 0 & m_{3}%
\end{array}%
\right) ,\hspace{0.5cm}\mbox{with}\hspace{0.5cm}V_{\nu }=\left( 
\begin{array}{ccc}
\cos \theta  & 0 & \sin \theta  \\ 
0 & 1 & 0 \\ 
-\sin \theta  & 0 & \cos \theta 
\end{array}%
\right) ,\hspace{0.5cm}\theta =\pm \frac{\pi }{4},  \label{Vnu}
\end{equation}%
where $\theta =+\pi /4$ and $\theta =-\pi /4$ correspond to normal (NH) and
inverted (IH) mass hierarchies, respectively. The masses for the light
active neutrinos, in the cases of normal (NH) and inverted (IH) mass
hierarchies, read: 
\begin{eqnarray}
\mbox{NH} &:&\theta =+\frac{\pi }{4}:\hspace{10mm}m_{\nu _{1}}=0,\hspace{10mm%
}m_{\nu _{2}}=B,\hspace{10mm}m_{\nu _{3}}=2A,  \label{mass-spectrum-Inverted}
\\[0.12in]
\mbox{IH} &:&\theta =-\frac{\pi }{4}:\hspace{10mm}m_{\nu _{1}}=2A,\hspace{8mm%
}m_{\nu _{2}}=B,\hspace{10mm}m_{\nu _{3}}=0.  \label{mass-spectrum-Normal}
\end{eqnarray}%
Besides that, the Pontecorvo-Maki-Nakagawa-Sakata (PMNS) leptonic mixing
matrix has the following form: 
\begin{equation}
U=V_{lL}^{\dag }P_{l}V_{\nu }=\left( 
\begin{array}{ccc}
\frac{e^{7i\phi }\cos \theta }{\sqrt{3}}-\frac{e^{2i\phi }\sin \theta }{%
\sqrt{3}} & \frac{e^{4i\phi }}{\sqrt{3}} & \frac{e^{2i\phi }\cos \theta }{%
\sqrt{3}}+\frac{e^{7i\phi }\sin \theta }{\sqrt{3}} \\ 
&  &  \\ 
\frac{e^{7i\phi }\cos \theta }{\sqrt{3}}-\frac{e^{2i\phi +\frac{2i\pi }{3}%
}\sin \theta }{\sqrt{3}} & \frac{e^{4i\phi -\frac{2i\pi }{3}}}{\sqrt{3}} & 
\frac{e^{2i\phi +\frac{2i\pi }{3}}\cos \theta }{\sqrt{3}}+\frac{e^{7i\phi
}\sin \theta }{\sqrt{3}} \\ 
&  &  \\ 
\frac{e^{7i\phi }\cos \theta }{\sqrt{3}}-\frac{e^{2i\phi -\frac{2i\pi }{3}%
}\sin \theta }{\sqrt{3}} & \frac{e^{4i\phi +\frac{2i\pi }{3}}}{\sqrt{3}} & 
\frac{e^{2i\phi -\frac{2i\pi }{3}}\cos \theta }{\sqrt{3}}+\frac{e^{7i\phi
}\sin \theta }{\sqrt{3}}%
\end{array}%
\right) .  \label{PMNS}
\end{equation}%
Note that while the PMNS leptonic mixing matrix only depends on a single
parameter $\phi $ (which arises from the charged lepton mass matrix), the
neutrino mass squared splittings are determined by two parameters, i.e., $A$
and $B$.

\quad Furthermore, we find that the lepton mixing angles are given by: 
\begin{eqnarray}
&&\sin ^{2}\theta _{12}=\frac{\left\vert U_{e2}\right\vert ^{2}}{%
1-\left\vert U_{e3}\right\vert ^{2}}=\frac{1}{2\mp \cos 5\phi },\hspace{20mm}
\label{theta-ij} \\[3mm]
&&\sin ^{2}\theta _{13}=\left\vert U_{e3}\right\vert ^{2}=\frac{1}{3}(1\pm
\cos 5\phi ), \\
&&\sin ^{2}\theta _{23}=\frac{\left\vert U_{\mu 3}\right\vert ^{2}}{%
1-\left\vert U_{e3}\right\vert ^{2}}=\frac{1}{2}\pm \frac{\sqrt{3}\sin 5\phi
)}{4\mp 2\cos 5\phi }.
\end{eqnarray}%
Then, from Eq. (\ref{PMNS})
, it follows that the limit $\phi =0$ and $\phi =\pi $ for the inverted and
normal neutrino mass hierarchies, respectively, correspond to the trimaximal
mixing, which predicts a vanishing reactor mixing angle. 
Let's note that the mixing angles for the lepton sector only depend on a
single parameter ($\phi $), while the neutrino mass squared splittings are
controlled by two parameters, i.e., $A$ and $B$.

\quad The Jarlskog invariant and the CP violating phase are given by \cite%
{PDG}: 
\begin{equation}
J=\func{Im}\left( U_{e1}U_{\mu 2}U_{e2}^{\ast }U_{\mu 1}^{\ast }\right) =-%
\frac{1}{6\sqrt{3}}\cos 2\theta ,\hspace{2cm}\sin \delta =\frac{8J}{\cos
\theta _{13}\sin 2\theta _{12}\sin 2\theta _{23}\sin 2\theta _{13}}.
\end{equation}%
From the relation $\theta =\pm \frac{\pi }{4}$, we predict $J=0$ and $\delta
=0$, which implies a vanishing leptonic Dirac CP violating phase.%

\quad 
In the following the three free effective parameters $\phi $, $A$ and $B$ of
the SM lepton sector of our model are adjusted to accommodate the
experimental values of 
three leptonic mixing parameters and two neutrino mass squared splittings,
shown in \mbox{Tables
\ref{NH}, \ref{IH}}, for the normal (NH) and inverted (IH) neutrino mass
hierarchies, respectively. The parameter $\phi $ is adjusted to reproduce
the experimental values of the leptonic mixing parameters $\sin ^{2}\theta
_{ij}$, whereas $A$ and $B$ for the normal (NH) and inverted (IH) neutrino
mass hierarchies read: 
\begin{eqnarray}
&&\mbox{NH}:\ m_{\nu _{1}}=0,\ \ \ m_{\nu _{2}}=B=\sqrt{\Delta m_{21}^{2}}%
\approx 9\mbox{meV},\ \ \ m_{\nu _{3}}=2A=\sqrt{\Delta m_{31}^{2}}\approx 50%
\mbox{meV};  \label{AB-Delta-IH} \\[0.12in]
&&\mbox{IH}\hspace{2mm}:\ m_{\nu _{2}}=B=\sqrt{\Delta m_{21}^{2}+\Delta
m_{13}^{2}}\approx 50\mbox{meV},\ \ \ \ \ m_{\nu _{1}}=2A=\sqrt{\Delta
m_{13}^{2}}\approx 49\mbox{meV},\ \ \ m_{\nu _{3}}=0,  \label{AB-Delta-NH}
\end{eqnarray}%
which follows from Eqs. (\ref{mass-spectrum-Normal}), (\ref%
{mass-spectrum-Inverted}) and the definition $\Delta
m_{ij}^{2}=m_{i}^{2}-m_{j}^{2}$. We take the best fit values of $\Delta
m_{ij}^{2}$ from Tables \ref{NH} and \ref{IH} for the normal and inverted
neutrino mass hierarchies, respectively.

\quad 
To reproduce the experimental values of the leptonic mixing parameters $\sin
^{2}\theta _{ij}$ given in Tables \ref{NH}, \ref{IH}, we vary the $\phi $
parameter, finding the following result: 
\begin{eqnarray}
&&\mbox{NH}\ :\ \phi =0.576\,\pi ,\ \ \ \sin ^{2}\theta _{12}\approx 0.34,\
\ \ \sin ^{2}\theta _{23}\approx 0.61,\ \ \ \sin ^{2}\theta _{13}\approx
0.0232;  \label{parameter-fit-IH} \\[0.12in]
&&\mbox{IH}\hspace{2.5mm}:\ \phi =\ \ 0.376\pi ,\ \ \ \ \ \sin ^{2}\theta
_{12}\approx 0.34,\ \ \ \sin ^{2}\theta _{23}\approx 0.61,\ \ \ \ \,\sin
^{2}\theta _{13}\approx 0.0238.  \label{parameter-fit-NH}
\end{eqnarray}

\quad Consequently, we find 
that $\sin ^{2}\theta _{13}$ is in excellent agreement with the experimental
data, for both normal and inverted neutrino mass hierarchies, whereas $\sin
^{2}\theta _{12}$ and $\sin ^{2}\theta _{23}$\ stay in the experimentally
allowed $2\sigma $ 
range. Thus, our predictions for the neutrino mass squared splittings and
leptonic mixing parameters, 
are in very good agreement with the experimental data on neutrino
oscillations, for both normal and inverted mass hierarchies. Furthermore,
another relevant prediction of our model is a vanishing leptonic Dirac CP
violating phase.
\begin{table}[tbh]
\begin{tabular}{|c|c|c|c|c|c|}
\hline
Parameter & $\Delta m_{21}^{2}$($10^{-5}$eV$^2$) & $\Delta m_{31}^{2}$($%
10^{-3}$eV$^2$) & $\left( \sin ^{2}\theta _{12}\right) _{\exp }$ & $\left(
\sin ^{2}\theta _{23}\right) _{\exp }$ & $\left( \sin ^{2}\theta
_{13}\right) _{\exp }$ \\ \hline
Best fit & $7.60$ & $2.48$ & $0.323$ & $0.567$ & $0.0234$ \\ \hline
$1\sigma $ range & $7.42-7.79$ & $2.41-2.53$ & $0.307-0.339$ & $0.439-0.599$
& $0.0214-0.0254$ \\ \hline
$2\sigma $ range & $7.26-7.99$ & $2.35-2.59$ & $0.292-0.357$ & $0.413-0.623$
& $0.0195-0.0274$ \\ \hline
$3\sigma $ range & $7.11-8.11$ & $2.30-2.65$ & $0.278-0.375$ & $0.392-0.643$
& $0.0183-0.0297$ \\ \hline
\end{tabular}%
\caption{Range for experimental values of neutrino mass squared splittings
and leptonic mixing parameters, taken from Ref. \protect\cite{Forero:2014bxa}%
, for the case of normal hierarchy.}
\label{NH}
\end{table}
\begin{table}[tbh]
\begin{tabular}{|c|c|c|c|c|c|}
\hline
Parameter & $\Delta m_{21}^{2}$($10^{-5}$eV$^{2}$) & $\Delta m_{13}^{2}$($%
10^{-3}$eV$^{2}$) & $\left( \sin ^{2}\theta _{12}\right) _{\exp }$ & $\left(
\sin ^{2}\theta _{23}\right) _{\exp }$ & $\left( \sin ^{2}\theta
_{13}\right) _{\exp }$ \\ \hline
Best fit & $7.60$ & $2.38$ & $0.323$ & $0.573$ & $0.0240$ \\ \hline
$1\sigma $ range & $7.42-7.79$ & $2.32-2.43$ & $0.307-0.339$ & $0.530-0.598$
& $0.0221-0.0259$ \\ \hline
$2\sigma $ range & $7.26-7.99$ & $2.26-2.48$ & $0.292-0.357$ & $0.432-0.621$
& $0.0202-0.0278$ \\ \hline
$3\sigma $ range & $7.11-8.11$ & $2.20-2.54$ & $0.278-0.375$ & $0.403-0.640$
& $0.0183-0.0297$ \\ \hline
\end{tabular}%
\caption{Range for experimental values of neutrino mass squared splittings
and leptonic mixing parameters, taken from Ref. \protect\cite{Forero:2014bxa}%
, for the case of inverted hierarchy.}
\label{IH}
\end{table}

\bigskip Now we determine the effective Majorana neutrino mass parameter,
which is proportional to the neutrinoless double beta ($0\nu \beta \beta $)
decay amplitude. The effective Majorana neutrino mass parameter is given by: 
\begin{equation}
m_{\beta \beta }=\left\vert \sum_{j}U_{ek}^{2}m_{\nu _{k}}\right\vert ,
\label{mee}
\end{equation}%
where $U_{ej}^{2}$ and $m_{\nu _{k}}$ are the PMNS mixing matrix elements
and the Majorana neutrino masses, respectively.

From Eqs. (\ref{PMNS}), (\ref{AB-Delta-IH}), (\ref{AB-Delta-NH}) and (\ref%
{mee}), we find that that the effective Majorana neutrino mass parameter,
for both Normal and Inverted hierarchies, takes the following values: 
\begin{equation}
m_{\beta \beta }=\left\{ 
\begin{array}{l}
3\ \mbox{meV}\ \ \ \ \ \ \ \mbox{for \ \ \ \ NH} \\ 
40\ \mbox{meV}\ \ \ \ \ \ \ \mbox{for \ \ \ \ IH} \\ 
\end{array}%
\right.  \label{eff-mass-pred}
\end{equation}%
%
%
%
%
%
%
%
%
%
%
%
%
%
%
%
Our obtained values $m_{\beta \beta }\approx 3\ \mbox{meV}$ and $m_{\beta
\beta }\approx 40\ \mbox{meV}$ for the effective Majorana neutrino mass
parameter, for normal and inverted hierarchies, respectively, are beyond the
reach of the present and forthcoming $0\nu \beta \beta $ decay experiments.
The current best upper bound on the effective neutrino mass is $m_{\beta
\beta }\leq 160$ meV, which corresponds to $T_{1/2}^{0\nu \beta \beta
}(^{136}\mathrm{Xe})\geq 1.6\times 10^{25}$ yr at 90\% C.L, as indicated by
the EXO-200 experiment \cite{Auger:2012ar}. This bound will be improved
within a not too far future. The GERDA \textquotedblleft
phase-II\textquotedblright experiment \cite{Abt:2004yk,Ackermann:2012xja} 
is expected to reach 
\mbox{$T^{0\nu\beta\beta}_{1/2}(^{76}{\rm Ge}) \geq
2\times 10^{26}$ yr}, which corresponds to $m_{\beta \beta }\leq 100$ meV. A
bolometric CUORE experiment, using ${}^{130}Te$ \cite{Alessandria:2011rc},
is currently under construction and has an estimated sensitivity of about $%
T_{1/2}^{0\nu \beta \beta }(^{130}\mathrm{Te})\sim 10^{26}$ yr, which
corresponds to \mbox{$m_{\beta\beta}\leq 50$ meV.} Furthermore, there are
proposals for ton-scale next-to-next generation $0\nu \beta \beta $
experiments with $^{136}$Xe \cite{KamLANDZen:2012aa,Albert:2014fya} and $%
^{76}$Ge \cite{Abt:2004yk,Guiseppe:2011me} claiming sensitivities over $%
T_{1/2}^{0\nu \beta \beta }\sim 10^{27}$ yr, which corresponds to $m_{\beta
\beta }\sim 12-30$ meV. For a recent review, see for example Ref. \cite%
{Bilenky:2014uka}. Consequently, as follows from Eq. (\ref{eff-mass-pred}),
our model predicts $T_{1/2}^{0\nu \beta \beta }$ at the level of
sensitivities of the next generation or next-to-next generation $0\nu \beta
\beta $ experiments.

Regarding the sterile neutrino sector, from Eqs. (\ref{Mnu2}) and (\ref{Mnu3}%
) we find that the sterile neutrino mass matrices are given by: 
\begin{eqnarray}
M_{\nu }^{\left( 2\right) } &=&-\frac{\sqrt{3}\left( h_{\chi }^{\left(
L\right) }\right) ^{2}v_{\chi }^{2}}{2h_{1N}v_{\xi }}\left( 
\begin{array}{ccc}
X & -Y & Y \\ 
-Y & X & Y \\ 
Y & Y & X%
\end{array}%
\right) ,\hspace{0.7cm}\hspace{0.7cm}X=\frac{x-1}{2x^{2}+x-1},\hspace{0.7cm}%
\hspace{0.7cm}Y=\frac{x}{2x^{2}+x-1}, \\
M_{\nu }^{\left( 3\right) } &=&h_{1N}\frac{v_{\xi }}{\sqrt{3}}\left( 
\begin{array}{ccc}
1 & -x & x \\ 
-x & 1 & x \\ 
x & x & 1%
\end{array}%
\right) ,\hspace{0.7cm}x=\frac{h_{2N}v_{S}}{h_{1N}v_{\xi }}.
\end{eqnarray}%
The sterile neutrino mass matrices $M_{\nu }^{\left( 2\right) }$ and $M_{\nu
}^{\left( 3\right) }$ are diagonalized by a rotation matrix $U_{R}=U_{\chi }$%
, according to: 
\begin{equation}
U_{R}^{T}M_{\nu }^{\left( k\right) }U_{R}=\left( 
\begin{array}{ccc}
M_{1}^{\left( \kappa \right) } & 0 & 0 \\ 
0 & M_{2}^{\left( \kappa \right) } & 0 \\ 
0 & 0 & M_{3}^{\left( \kappa \right) }%
\end{array}%
\right) ,\hspace{0.5cm}U_{R}=U_{\chi }=\left( 
\begin{array}{ccc}
-\frac{1}{\sqrt{3}} & -\frac{1}{\sqrt{2}} & -\frac{1}{\sqrt{6}} \\ 
-\frac{1}{\sqrt{3}} & \frac{1}{\sqrt{2}} & -\frac{1}{\sqrt{6}} \\ 
\frac{1}{\sqrt{3}} & 0 & -\sqrt{\frac{2}{3}}%
\end{array}%
\right) ,\hspace{0.5cm}k=2,3,
\end{equation}%
where the sterile neutrino masses are given by: 
\begin{eqnarray}
M_{1}^{\left( 2\right) } &=&\left\vert \frac{\sqrt{3}\left( h_{\chi
}^{\left( L\right) }\right) ^{2}v_{\chi }^{2}}{2h_{1N}v_{\xi }}\left(
X-2Y\right) \right\vert ,\hspace{0.5cm}\hspace{0.5cm}M_{2}^{\left( 2\right)
}=M_{3}^{\left( 2\right) }=\left\vert \frac{\sqrt{3}\left( h_{\chi }^{\left(
L\right) }\right) ^{2}v_{\chi }^{2}}{2h_{1N}v_{\xi }}\left( X+Y\right)
\right\vert , \\
M_{1}^{\left( 3\right) } &=&\left\vert h_{1N}\frac{v_{\xi }}{\sqrt{3}}\left(
1-2x\right) \right\vert ,\hspace{0.5cm}\hspace{0.5cm}M_{2}^{\left( 3\right)
}=M_{3}^{\left( 3\right) }=\left\vert h_{1N}\frac{v_{\xi }}{\sqrt{3}}\left(
1+x\right) \right\vert ,
\end{eqnarray}%
which implies that the heavy and very heavy sterile neutrino spectrum
includes two degenerates heavy neutrino and one light neutrino states.

\quad Furthermore, as follows from Eq. (\ref{U}) and the relation $%
\left\vert \left( B_{2}\right) _{ij}\right\vert \sim 10^{-5}<<\left\vert
\left( B_{1}\right) _{ij}\right\vert \sim 10^{-3}$ ($i,j=1,2,3$) given by
Eq. (\ref{Binequality}), 
we can connect the neutrino fields $\nu _{L}=\left( \nu _{1L},\nu _{2L},\nu
_{3L}\right) ^{T}$, $\nu _{R}^{C}=\left( \nu _{1R}^{C},\nu _{2R}^{C},\nu
_{3R}^{C}\right) $ and $N_{R}^{C}=\left(N_{1R}^{C},N_{2R}^{C},N_{3R}^{C}%
\right) $ with the neutrino mass eigenstates by the following approximate
relations: 
\begin{equation}
\left( 
\begin{array}{c}
\nu _{L} \\ 
\nu _{R}^{C} \\ 
N_{R}^{C}%
\end{array}%
\right)\simeq\left( 
\begin{array}{c}
V_{\nu }\xi _{L}^{\left( 1\right) } \\ 
U_{R}\xi _{L}^{\left( 2\right) }+B_{1}U_{R}\xi _{L}^{\left( 3\right) } \\ 
U_{R}\xi _{L}^{\left( 3\right) }-B_{1}^{\dagger }U_{R}\xi _{L}^{\left(
2\right) }%
\end{array}%
\right) ,\hspace{0.5cm}\hspace{0.5cm}\hspace{0.5cm}\hspace{0.5cm}\xi
_{L}^{\left( j\right) }=\left( 
\begin{array}{c}
\xi _{1L}^{\left( j\right) } \\ 
\xi _{2L}^{\left( j\right) } \\ 
\xi _{3L}^{\left( j\right) }%
\end{array}%
\right) ,\hspace{0.5cm}\hspace{0.5cm}\hspace{0.5cm}\hspace{0.5cm}j=1,2,3.
\end{equation}
where $\xi _{kL}^{\left( 1\right) }$, $\xi _{kL}^{\left( 2\right) }$ and $%
\xi _{kL}^{\left( 3\right) }$ ($k=1,2,3$) are the light active, heavy
sterile and very heavy sterile neutrinos, respectively. As previously
mentioned, the heavy sterile neutrinos have MeV scale masses and thus
correspond to dark matter candidates. Furthermore, we assume that the
lightest of the very heavy sterile neutrinos, i.e., $\xi
_{1L}^{\left(3\right) }$ has a TeV scale mass and thus corresponds to a
candidate for detection at the LHC.

\section{Quark masses and mixing.}

\label{quarkmassesandmixing} From Eq. (\ref{Yukawaterms}) and taking into
account that the VEV pattern of the $SU\left( 3\right) _{L}$ singlet scalar
fields is described by Eq. (\ref{VEV}), with the nonvanishing VEVs set to be
equal to $\lambda\Lambda$ (being $\Lambda $ the cutoff of our model)\ as
shown in Eq. (\ref{VEVsinglets}), it follows that the SM quark mass matrices
have the form: 
\begin{equation}
M_{U}=\left( 
\begin{array}{ccc}
a_{11}^{\left( U\right) }\lambda ^{4} & a_{12}^{\left( U\right) }\lambda ^{3}
& a_{13}^{\left( U\right) }\lambda ^{2} \\ 
a_{21}^{\left( U\right) }\lambda ^{3} & a_{22}^{\left( U\right) }\lambda ^{2}
& a_{23}^{\left( U\right) }\lambda \\ 
a_{31}^{\left( U\right) }\lambda ^{2} & a_{32}^{\left( U\right) }\lambda & 
a_{33}^{\left( U\right) }%
\end{array}%
\right) \frac{v}{\sqrt{2}},\hspace{1cm}\hspace{1cm}M_{D}=\left( 
\begin{array}{ccc}
a_{11}^{\left( D\right) }\lambda ^{7} & a_{12}^{\left( D\right) }\lambda ^{6}
& a_{13}^{\left( D\right) }\lambda ^{5} \\ 
a_{21}^{\left( D\right) }\lambda ^{6} & a_{22}^{\left( D\right) }\lambda ^{5}
& a_{23}^{\left( D\right) }\lambda ^{4} \\ 
a_{31}^{\left( D\right) }\lambda ^{5} & a_{32}^{\left( D\right) }\lambda ^{4}
& a_{33}^{\left( U\right) }\lambda ^{3}%
\end{array}%
\right) \frac{v}{\sqrt{2}},  \label{Mq}
\end{equation}%
where $\lambda =0.225$ is one of the Wolfenstein parameters, $v=246$ GeV the
scale of electroweak symmetry breaking and $a_{ij}^{\left( U,D\right) }$ ($%
i,j=1,2,3$) are $\mathcal{O}(1)$ parameters given by the following
relations: 
\begin{eqnarray}
a_{nj}^{\left( U\right) } &=&y_{nj}^{\left( U\right) }\frac{v_{\rho }}{v}%
e^{i\left( 6-j-n\right) \phi },\hspace{1cm}a_{3j}^{\left( U\right)
}=y_{3j}^{\left( U\right) }\frac{v_{\eta }}{v}e^{i\left( 3-j\right) \phi },%
\hspace{1cm}  \notag \\
a_{nj}^{\left( D\right) } &=&y_{nj}^{\left( D\right) }\frac{v_{\eta }}{v}%
e^{i\left( 6-j-n\right) \phi },\hspace{1cm}a_{3j}^{\left( D\right)
}=y_{3j}^{\left( D\right) }\frac{v_{\rho }}{v}e^{i\left( 3-j\right) \phi },%
\hspace{1cm}n=1,2,\hspace{1cm}j=1,2,3.  \label{aq}
\end{eqnarray}

Furthermore, the exotic quark masses read: 
\begin{equation}
m_{T}=y^{\left( T\right) }\frac{v_{\chi }}{\sqrt{2}},\hspace{1cm}%
m_{J^{1}}=y_{1}^{\left( J\right) }\frac{v_{\chi }}{\sqrt{2}}=\frac{%
y_{1}^{\left( J\right) }}{y^{\left( T\right) }}m_{T},\hspace{1cm}%
m_{J^{2}}=y_{2}^{\left( J\right) }\frac{v_{\chi }}{\sqrt{2}}=\frac{%
y_{2}^{\left( J\right) }}{y^{\left( T\right) }}m_{T}.  \label{mexotics}
\end{equation}

Since the charged fermion mass and quark mixing pattern arises from the
breaking of the $Z_{2}\otimes Z_{3}\otimes Z_{14}$ discrete group, we
asssume an approximate universality in the dimensionless SM quark Yukawa
couplings, as follows: 
\begin{eqnarray}
a_{11}^{\left( U\right) } &=&a_{1}^{\left( U\right) }e^{i\phi _{1}},\hspace{%
1cm}a_{22}^{\left( U\right) }=a_{2}^{\left( U\right) },\hspace{1cm}%
a_{33}^{\left( U\right) }=a_{3}^{\left( U\right) },  \notag \\
a_{12}^{\left( U\right) } &=&a_{1}^{\left( U\right) }\left( 1-\frac{\lambda
^{2}}{2}\right) ^{-\frac{3}{2}}e^{i\phi _{2}},\hspace{1cm}a_{13}^{\left(
U\right) }=a_{2}^{\left( U\right) }\left( 1-\frac{\lambda ^{2}}{2}\right) ^{-%
\frac{3}{2}}e^{i\phi _{2}},\hspace{1cm}a_{23}^{\left( U\right) }=\left\vert
a_{13}^{\left( U\right) }\right\vert \left( 1-\frac{\lambda ^{2}}{2}\right)
^{-\frac{3}{2}},  \notag \\
a_{11}^{\left( D\right) } &=&a_{22}^{\left( D\right) }\left( 1-\frac{\lambda
^{2}}{2}\right) ^{-2},\hspace{1cm}a_{23}^{\left( D\right) }=a_{33}^{\left(
D\right) }\left( 1-\frac{\lambda ^{2}}{2}\right) ^{-\frac{1}{2}},\hspace{1cm}%
a_{ij}^{\left( U,D\right) }=a_{ji}^{\left( U,D\right) },\hspace{1cm}%
i,j=1,2,3,  \label{aQ}
\end{eqnarray}

It is noteworthy that exact universality in the dimensionless quark Yukawa
couplings predicts massless up, down, strange and charm quarks, in
contradiction with the experimental data on quark masses. Consequently, to
generate these masses, the universality in the quark Yukawa couplings has to
be broken. 
Besides that, for simplicity, we assume that the complex phase responsible
for CP violation in the quark sector only arises from up type quark Yukawa
terms, as indicated by Eq. (\ref{aQ}). In addition, for the sake of
simplicity, we fix $a_{3}^{\left( U\right) }=1$, which is suggested by
naturalness arguments. Therefore, the SM quark mass matrices are given by: 
\begin{equation}
M_{U}=\left( 
\begin{array}{ccc}
a_{1}^{\left( U\right) }\lambda ^{4}e^{i\phi _{1}} & a_{1}^{\left( U\right)
}\left( 1-\frac{\lambda ^{2}}{2}\right) ^{-\frac{3}{2}}\lambda ^{3}e^{i\phi
_{2}} & a_{2}^{\left( U\right) }\left( 1-\frac{\lambda ^{2}}{2}\right) ^{-%
\frac{3}{2}}\lambda ^{2}e^{i\phi _{2}} \\ 
a_{1}^{\left( U\right) }\left( 1-\frac{\lambda ^{2}}{2}\right) ^{-\frac{3}{2}%
}\lambda ^{3}e^{i\phi _{2}} & a_{2}^{\left( U\right) }\lambda ^{2} & 
a_{2}^{\left( U\right) }\left( 1-\frac{\lambda ^{2}}{2}\right) ^{-3}\lambda
\\ 
a_{2}^{\left( U\right) }\left( 1-\frac{\lambda ^{2}}{2}\right) ^{-\frac{3}{2}%
}\lambda ^{2}e^{i\phi _{2}} & a_{2}^{\left( U\right) }\left( 1-\frac{\lambda
^{2}}{2}\right) ^{-3}\lambda & a_{3}^{\left( U\right) }%
\end{array}%
\right) \frac{v}{\sqrt{2}},
\end{equation}

\begin{equation}
M_{D}=\left( 
\begin{array}{ccc}
a_{22}^{\left( D\right) }\left( 1-\frac{\lambda ^{2}}{2}\right) ^{-2}\lambda
^{7} & a_{12}^{\left( D\right) }\lambda ^{6} & a_{13}^{\left( D\right)
}\lambda ^{5} \\ 
a_{12}^{\left( D\right) }\lambda ^{6} & a_{22}^{\left( D\right) }\lambda ^{5}
& a_{33}^{\left( D\right) }\left( 1-\frac{\lambda ^{2}}{2}\right) ^{-\frac{1%
}{2}}\lambda ^{4} \\ 
a_{13}^{\left( D\right) }\lambda ^{5} & a_{33}^{\left( D\right) }\left( 1-%
\frac{\lambda ^{2}}{2}\right) ^{-\frac{1}{2}}\lambda ^{4} & a_{33}^{\left(
D\right) }\lambda ^{3}%
\end{array}%
\right) \frac{v}{\sqrt{2}}
\end{equation}

\quad Let's recall that the quark sector has 10 effective parameters, i.e, $%
\lambda $, $a_{3}^{\left( U\right) }$, $a_{1}^{\left( U\right) }$, $%
a_{2}^{\left( U\right) }$, $a_{22}^{\left( D\right) }$, $a_{12}^{\left(
D\right) }$, $a_{13}^{\left( D\right) }$, $a_{33}^{\left( D\right) }$ and
the phases $\phi _{1}$ and $\phi _{2}$ to describe the quark mass and mixing
pattern, which is determined by 10 observables. Nevertheless, not all these
effective parameters are free since the parameters $\lambda $ and $%
a_{3}^{\left( U\right) }$ are fixed while the remaining 8 parameters are
adjusted to reproduce the physical observables in the quark sector, i.e., 6
quark masses and 4 quark mixing parameters. The results shown in Table \ref%
{Observables} correspond to the following best-fit values: 
\begin{eqnarray}
a_{1}^{\left( U\right) } &\simeq &0.64,\hspace{1cm}a_{2}^{\left( U\right)
}\simeq 0.77,\hspace{1cm}\phi _{1}\simeq -9.03^{\circ },\hspace{1cm}\phi
_{2}\simeq -4.53^{\circ },  \notag \\
a_{22}^{\left( D\right) } &\simeq &2.03,\hspace{1cm}a_{12}^{\left( D\right)
}\simeq 1.75,\hspace{1cm}a_{13}^{\left( D\right) }\simeq 1.15,\hspace{1cm}%
a_{33}^{\left( D\right) }\simeq 1.40.
\end{eqnarray}

\begin{table}[tbh]
\begin{center}
\begin{tabular}{c|l|l}
\hline\hline
Observable & Model value & Experimental value \\ \hline
$m_{u}(MeV)$ & \quad $1.59$ & \quad $1.45_{-0.45}^{+0.56}$ \\ \hline
$m_{c}(MeV)$ & \quad $673$ & \quad $635\pm 86$ \\ \hline
$m_{t}(GeV)$ & \quad $180$ & \quad $172.1\pm 0.6\pm 0.9$ \\ \hline
$m_{d}(MeV)$ & \quad $2.9$ & \quad $2.9_{-0.4}^{+0.5}$ \\ \hline
$m_{s}(MeV)$ & \quad $59.7$ & \quad $57.7_{-15.7}^{+16.8}$ \\ \hline
$m_{b}(GeV)$ & \quad $2.98$ & \quad $2.82_{-0.04}^{+0.09}$ \\ \hline
$\bigl|V_{ud}\bigr|$ & \quad $0.975$ & \quad $0.97427\pm 0.00015$ \\ \hline
$\bigl|V_{us}\bigr|$ & \quad $0.224$ & \quad $0.22534\pm 0.00065$ \\ \hline
$\bigl|V_{ub}\bigr|$ & \quad $0.0036$ & \quad $0.00351_{-0.00014}^{+0.00015}$
\\ \hline
$\bigl|V_{cd}\bigr|$ & \quad $0.224$ & \quad $0.22520\pm 0.00065$ \\ \hline
$\bigl|V_{cs}\bigr|$ & \quad $0.9736$ & \quad $0.97344\pm 0.00016$ \\ \hline
$\bigl|V_{cb}\bigr|$ & \quad $0.0433$ & \quad $0.0412_{-0.0005}^{+0.0011}$
\\ \hline
$\bigl|V_{td}\bigr|$ & \quad $0.00853$ & \quad $%
0.00867_{-0.00031}^{+0.00029} $ \\ \hline
$\bigl|V_{ts}\bigr|$ & \quad $0.0426$ & \quad $0.0404_{-0.0005}^{+0.0011}$
\\ \hline
$\bigl|V_{tb}\bigr|$ & \quad $0.999057$ & \quad $%
0.999146_{-0.000046}^{+0.000021}$ \\ \hline
$J$ & \quad $2.98\times 10^{-5}$ & \quad $(2.96_{-0.16}^{+0.20})\times
10^{-5}$ \\ \hline
$\delta $ & \quad $61^{\circ }$ & \quad $68^{\circ }$ \\ \hline\hline
\end{tabular}%
\end{center}
\caption{Model and experimental values of the quark masses and CKM
parameters.}
\label{Observables}
\end{table}

The obtained and experimental values of the quark masses, CKM matrix
elements, Jarlskog invariant $J$ and CP violating phase $\delta $ are
reported in Table \ref{Observables}. We use the experimental values of the
quark masses at the $M_{Z}$ scale, from Ref. \cite{Bora:2012tx} (which are
similar to those in \cite{Xing:2007fb}), whereas we use the experimental
values of the CKM parameters from Ref. \cite{PDG}. The obtained values of
the quark masses and CKM parameters are in excellent agreement with the
experimental data, as indicated by Table \ref{Observables}. 

\section{Conclusions}

\label{conclusions}

In this paper we present an extension of the minimal $331$ model with $\beta
=-\frac{1}{\sqrt{3}}$, based on the extended $SU(3)_{C}\otimes
SU(3)_{L}\otimes U(1)_{X}\otimes T_{7}\otimes Z_{2}\otimes Z_{3}\otimes
Z_{14}$ symmetry. Our economical $T_{7}$ flavor 331 model, which at low
energies reduces to the minimal 331 model with $\beta =-\frac{1}{\sqrt{3}}$,
is compatible with the experimental data on fermion masses and mixing. The
model has in total 16 effective free parameters, which are fitted to
reproduce the experimental values of the 18 physical observables in the
quark and lepton sectors. The $T_{7}$ and $Z_{3}$ symmetries reduce the
number of model parameters. In particular, the $Z_{3}$ symmetry determines
the allowed entries of the neutrino mass matrix and decouples the SM quarks
from the exotic quarks. The $Z_{2}$ symmetry generates the hierarchy between
SM up and SM down type quark masses. We assumed that the $SU(3)_{L}$ scalar
singlets having Yukawa interactions with the right handed Majorana neutrinos
acquire VEVs at very high scale, then providing very large masses to these
Majorana neutrinos, and thus giving rise to a double seesaw mechanism of
active neutrino masses. 
Consequently, the neutrino spectrum includes very light active neutrinos as
well as heavy and very heavy sterile neutrinos. We find that the heavy and
very heavy sterile neutrinos have masses at the $\sim $ MeV and $\sim $ TeV
scales, respectively. Thus, the MeV scale sterile neutrinos of our model
correspond to dark matter candidates. The smallness of the active neutrino
masses is attributed to their scaling with inverse powers of the high energy
cutoff $\Lambda $ as well as well as by their quadratic dependence on the
very small VEV of the $Z_{2}\otimes Z_{3}\otimes Z_{14}$ neutral, $SU(3)_{L}$
singlet and $T_{7}$ antitriplet scalar field $\zeta $. The observed
hierarchy of charged fermion masses and quark mixing matrix elements arises
from the breaking of the $Z_{2}\otimes Z_{3}\otimes Z_{14}$ discrete group
at a very high energy. The tau, muon and electron masses arise from
effective seven, nine and twelve dimensional Yukawa operators, respectively.
We find for the scale of these operators the estimate $\Lambda \sim 10^{5}$
TeV. 
The complex phase responsible for CP violation in the quark sector has been
assumed to come from up-type quark Yukawa terms. The model predicts an
effective Majorana neutrino mass, relevant for neutrinoless double beta
decay, with values $m_{\beta \beta }=$ 3 and 40 meV, for the normal and the
inverted neutrino spectrum, respectively. In the latter case our prediction
is within the declared reach of the next generation bolometric CUORE
experiment \cite{Alessandria:2011rc} or, more realistically, of the
next-to-next generation tone-scale $0\nu \beta \beta $-decay experiments.
Furthermore, a vanishing leptonic Dirac CP violating phase is predicted in
our model.

\section*{Acknowledgments}

We are very grateful to Fredy Ochoa for careful reading of the manuscript
and for valuable discussions. A.E.C.H was supported by Fondecyt (Chile),
Grant No. 11130115 and by DGIP internal Grant No. 111458. R.M. was supported
by El Patrimonio Aut\'{o}nomo Fondo Nacional de Financiamiento para la
Ciencia, la Tecnolog\'{\i}a y la Innovaci\'{o}n Fransisco Jos\'{e} de Caldas
programme of COLCIENCIAS in Colombia.

\appendix

\section{The product rules for $T_{7}$ \label{A}}

\label{ap1} The group $T_{7}$, which is a subgroup of $SU(3)$ and $\Delta
(3N^{2})$ with $N=7$, has $21$ elements, is isomorphic to $Z_{7}\rtimes
Z_{3} $ and contains five irreducible representations, i.e., one triplet $%
\mathbf{3}$, one antitriplet $\mathbf{\bar{3}}$ and three singlets $\mathbf{1%
}_{0}$, $\mathbf{1}_{1}$ and $\mathbf{1}_{2}$ \cite{Ishimori:2010au}. The
discrete group $T_{7}$ is the minimal non-Abelian discrete group having a
complex triplet. The triplet and antitriplet irreducible representations are
defined as \cite{Ishimori:2010au}: 
\begin{equation}
\mathbf{3}\equiv \3tvec{x_{1}}{x_{2}}{x_{4}},\quad \mathbf{\bar{3}}\equiv \3%
tvec{x_{-1}}{x_{-2}}{x_{-4}}=\3tvec{x_{6}}{x_{5}}{x_{3}}.
\end{equation}%
The product rules for triplet and antitriplet tensor irreducible
representations are given by: 
\begin{eqnarray}
\3tvec{x_{1}}{x_{2}}{x_{4}}_{\mathbf{3}}\otimes \3tvec{y_{1}}{y_{2}}{y_{4}}_{%
\mathbf{3}} &=&\3tvec{x_{2}y_{4}}{x_{4}y_{1}}{x_{1}y_{2}}_{\mathbf{\bar{3}}%
}\oplus \3tvec{x_{4}y_{2}}{x_{1}y_{4}}{x_{2}y_{1}}_{\mathbf{\bar{3}}}\oplus %
\3tvec{x_{4}y_{4}}{x_{1}y_{1}}{x_{2}y_{2}}_{\mathbf{3}}, \\
\3tvec{x_{6}}{x_{5}}{x_{3}}_{\mathbf{\bar{3}}}\otimes \3tvec{y_{6}}{y_{5}}{%
y_{3}}_{\mathbf{\bar{3}}} &=&\3tvec{x_{5}y_{3}}{x_{3}y_{6}}{x_{6}y_{5}}_{%
\mathbf{3}}\oplus \3tvec{x_{3}y_{5}}{x_{6}y_{3}}{x_{5}y_{6}}_{\mathbf{3}%
}\oplus \3tvec{x_{3}y_{3}}{x_{6}y_{6}}{x_{5}y_{5}}_{\mathbf{\bar{3}}}, \\
\3tvec{x_{1}}{x_{2}}{x_{4}}_{\mathbf{3}}\otimes \3tvec{y_{6}}{y_{5}}{y_{3}}_{%
\mathbf{\bar{3}}} &=&\3tvec{x_{2}y_{6}}{x_{4}y_{5}}{x_{1}y_{3}}_{\mathbf{3}%
}\oplus \3tvec{x_{1}y_{5}}{x_{2}y_{3}}{x_{4}y_{6}}_{\mathbf{\bar{3}}}  \notag
\\
&&\oplus \sum_{k=0,1,2}(x_{1}y_{6}+\omega ^{k}x_{2}y_{5}+\omega
^{2k}x_{4}y_{3})_{\mathbf{1}_{k}}.
\end{eqnarray}%
%
%
%
%
%
%
%
%
%
%
%
%
%
%
%
%
%
%
%
%
%
%
%
%
%
%
%
%
%
%
%
%
%
%
%
%
%
%
%
%
%
%
%
%
%
%
%
%
%
%
%
%
%
%
%
%
Whereas the tensor products between singlets are: 
\begin{eqnarray}
(x)_{\mathbf{1}_{0}}(y)_{\mathbf{1}_{0}}&=&(x)_{\mathbf{1}_{1}}(y)_{\mathbf{1%
}_{2}}=(x)_{\mathbf{1}_{2}}(y)_{\mathbf{1}_{1}}=(xy)_{\mathbf{1}_{0}},~ 
\notag \\
(x)_{\mathbf{1}_{1}}(y)_{\mathbf{1}_{1}}&=&(xy)_{\mathbf{1}_{2}},~  \notag \\
(x)_{\mathbf{1}_{2}}(y)_{\mathbf{1}_{2}}&=&(xy)_{\mathbf{1}_{1}}.
\end{eqnarray}%
%
%
%
%
%
%
%
%
%
%
%
%
%
%
%
%
%
%
%
%
%
%
%
%
%
%
The product rules between triplets and singlets satisfy the relations: 
\begin{equation}
(y)_{\mathbf{1}_{k}}\otimes \3tvec{x_{1(6)}}{x_{2(5)}}{x_{4(3)}}_{\mathbf{3(%
\bar{3})}}=\3tvec{yx_{1(6)}}{yx_{2(5)}}{yx_{4(3)}}_{\mathbf{3(\bar{3})}}.
\end{equation}%
where $\omega =e^{i\frac{2\pi }{3}}$. The representation $\mathbf{1}_{0}$ is
trivial, while the non-trivial $\mathbf{1}_{1}$ and $\mathbf{1}_{2}$ are
complex conjugate to each other. Some reviews of discrete symmetries in
particle physics are found in Refs. \cite%
{King:2013eh,Altarelli:2010gt,Ishimori:2010au,Discret-Group-Review}.

\end{document}

%% file: convention.tex

\def\2tvec#1#2{
\left(
\begin{array}{c}
#1  \\
#2  \\   
\end{array}
\right)}

\def\mat2#1#2#3#4{
\left(
\begin{array}{cc}
#1 & #2 \\
#3 & #4 \\
\end{array}
\right)
}

\def\Mat3#1#2#3#4#5#6#7#8#9{
\left(
\begin{array}{ccc}
#1 & #2 & #3 \\
#4 & #5 & #6 \\
#7 & #8 & #9 \\
\end{array}
\right)
}

\def\3tvec#1#2#3{
\left(
\begin{array}{c}
#1  \\
#2  \\   
#3  \\
\end{array}
\right)}

\def\4tvec#1#2#3#4{
\left(
\begin{array}{c}
#1  \\
#2  \\   
#3  \\
#4  \\
\end{array}
\right)}

\def\5tvec#1#2#3#4#5{
\left(
\begin{array}{c}
#1  \\
#2  \\
#3  \\
#4  \\
#5  \\
\end{array}
\right)}

\def\L{\left}
\def\R{\right}

\def\pl{\partial}

\def\lra{\leftrightarrow}
